\documentclass[journal]{IEEEtran}

\ifCLASSINFOpdf
\usepackage[pdftex]{graphicx}
\DeclareGraphicsExtensions{.pdf,.jpeg,.png}
\else
\usepackage[dvips]{graphicx}
\fi
\usepackage{subfigure}
\usepackage{epstopdf}
\usepackage[cmex10]{amsmath}
\usepackage{amssymb}
\usepackage{wasysym}
\usepackage{algorithm}
\usepackage{algorithmic}
\usepackage{array}
\usepackage{cite}
\usepackage{color}
\usepackage{url}
\usepackage{bm}
\usepackage{enumerate}
\usepackage{epsfig,latexsym}%amsmath
\usepackage{graphicx}

\usepackage{cuted}
\usepackage{lipsum}
\begin{document}
\title{Simultaneous Beam Training and Target Sensing in ISAC Systems with RIS}
\author{Kangjian~Chen,~\IEEEmembership{Student~Member,~IEEE}, Chenhao~Qi,~\IEEEmembership{Senior~Member,~IEEE}, \\ Octavia A. Dobre,~\IEEEmembership{Fellow,~IEEE}  and Geoffrey Ye Li,~\IEEEmembership{Fellow,~IEEE}
	\thanks{This work is supported in part by the National Natural Science Foundation of China under Grants U22B2007 and 62071116. The work of Kangjian Chen is supported in part by the Postgraduate Research\&Practice Innovation Program of Jiangsu Province under Grant KYCX23\_0262. The work of Octavia A. Dobre is supported in part by the Natural Sciences and Engineering Research Council of Canada (NSERC) through its Discovery program. This paper has been presented in part at the 2023 IEEE International Conference on Communications (ICC), Rome, Italy, May 2023~\cite{icc23CKJ}. (\textit{Corresponding author: Chenhao~Qi}) }
	\thanks{Kangjian~Chen and Chenhao~Qi are with the School of Information Science and Engineering, Southeast University, Nanjing 210096, China (e-mail: qch@seu.edu.cn).}
	\thanks{Octavia A. Dobre is with the Faculty of Engineering and Applied Science, Memorial University, St. John’s, NL A1C 5S7, Canada (e-mail: odobre@mun.ca).}
	\thanks{Geoffrey Ye Li is with the Department of Electrical and Electronic Engineering, Imperial College London, SW7 2AZ London, U.K. (e-mail: geoffrey.li@imperial.ac.uk).}
}


\markboth{Accepted by IEEE Transactions on Wireless Communications}
{}

\maketitle
	
\begin{abstract}
This paper investigates an integrated sensing and communication (ISAC) system with reconfigurable intelligent surface (RIS). Our simultaneous beam training and target sensing (SBTTS) scheme enables the base station to perform beam training with the user terminals (UTs) and the RIS, and simultaneously to sense the targets. Based on our findings, the energy of the echoes from the RIS is accumulated in the angle-delay domain while that from the targets is accumulated in the Doppler-delay domain. The SBTTS scheme can distinguish the RIS from the targets with the mixed echoes from the RIS and the targets. Then we propose a positioning and array orientation estimation (PAOE) scheme for both the line-of-sight channels and the non-line-of-sight channels based on the beam training results of SBTTS by developing a low-complexity two-dimensional fast search algorithm. Based on the SBTTS and PAOE schemes, we further compute the angle-of-arrival and angle-of-departure for the channels between the RIS and the UTs by exploiting the geometry relationship to accomplish the beam alignment of the ISAC system. Simulation results verify the effectiveness of the proposed schemes.

\end{abstract}

\section{Introduction}
Wireless communication and radar sensing systems have been developed separately for several decades. With the evolution of technologies, these systems share similar aspects including antenna array, carrier frequency, radio frequency (RF) chains, and signal processing algorithms~\cite{JSTSP21ZhangAndrew, CST22_ZhangAndrew,CST21LN,Tcom20Liufan}. On the other hand, many emerging applications, such as autonomous driving and unmanned aerial vehicles, have strong demands for both wireless communication and radar sensing. Therefore, the similar aspects and application requirements have inspired the integrated sensing and communication (ISAC)~\cite{CST22LiuAn}.

% For example, in~\cite{TSP16LiBo}, interference power at the radar receiver is minimized subject to a certain data transmission rate by carefully designing the communication transmit covariance matrix. In~\cite{JSTSP18Zhengle}, the radar waveform is first estimated and then canceled prior to demodulation at the communication receiver. 

%In \cite{JSAC22Niukai} and \cite{JSAC22Qinshi}, the multi-station collaborative localization is investigated, where the kinematic parameters are extracted from the echoes of communication signals. In \cite{TWC20Yuanweijie} and \cite{Yuan2021Integrated}, radar sensing is exploited to assist the beam alignment and precoding of wireless communications, respectively. 

Earlier works on ISAC explore the coexistence of communication and sensing and focus on mitigating the interference between the communication and the sensing systems while maintaining their functionalities~\cite{TSP16LiBo,JSTSP18Zhengle}. However, the wireless communication and radar sensing systems still operate independently, which usually needs two sets of hardware resources. To improve the hardware and the spectral efficiency, recent efforts integrate the communication and sensing into a single system so that they can efficiently share space, time and frequency resources~\cite{CL22Chenhao,JSAC22YeZhifan}. Besides the system-level integration, the complementary functionalities of communication and sensing are recently exploited to improve the whole system performance~\cite{JSAC22Niukai,JSAC22Qinshi,TWC20Yuanweijie}.

%Existing works on ISAC mainly focuses on three topics, including the coexistence of communication and sensing, the dual-functional designs of communication and sensing and the mutualism of communication and sensing~\cite{JSAC22Liufan}. The first topic focuses on mitigating the interference between the communication systems and the sensing systems. The second topic consists of a variety of subproblems, such as dual-functional beamforming design, dual-functional waveform design and so on. The third topic focuses on exploiting the sensing ability to improve the communication performance or extracting pivotal information from communication signals to assist sensing. 

Recently, reconfigurable intelligent surface (RIS) has been proposed as an effective way to  expand the signal coverage~\cite{TWC21Wangwei,WCL20Ningboyu}, where channel state information (CSI) acquisition is a challenge. Generally, the RIS without an active transmitter or receiver can only passively reflect the incident signal. As a result, when a base station (BS) serves multiple user terminals (UTs) with the assistance of RISs, estimating the cascaded BS-RIS-UT channel usually requires a large overhead of pilots. Therefore, many works try to reduce the pilot overhead in various scenarios~\cite{TWC20WZR,WC22WPL,TVT21Ningboyu}. In~\cite{TWC20WZR}, the overhead for multiuser channel estimation is reduced by exploiting the property that all the UTs share the common channel from the BS to the RIS. The multi-directional beam training method in \cite{WC22WPL} can reduce the training overhead for the BS-RIS-UT channel. In the RIS-aided terahertz multiuser systems in~\cite{TVT21Ningboyu}, two dedicated hierarchical codebooks are designed to reduce the searching overhead while maintaining the robustness against the noise. 

The RIS can also be introduced into ISAC to design the RIS-assisted ISAC system~\cite{TVT22Liuyu,IEEESysJ21JZ,JSTSP22Liurang,JSAC22Heyinghui}. In~\cite{TVT22Liuyu}, a deep-learning framework is adopted to deal with the channel estimation problem in a RIS-assisted ISAC system. According to~\cite{IEEESysJ21JZ}, the RIS can improve the radar detection performance while maintaining the communication capability. In~\cite{JSTSP22Liurang}, the reflection coefficients of the RIS elements are optimized to simultaneously improve the sensing and the communication performance. In~\cite{JSAC22Heyinghui}, two RISs are deployed for effective communication signal enhancement and  mutual interference mitigation. 

%Note that the BS of the emerging ISAC system is endowed with the ability of sensing, which brings tremendous vitality to the conventional communication systems. 
Different from the existing works that estimate  CSI solely by conventional methods in the RIS-assisted ISAC system, we will facilitate the CSI acquisition of the RIS-assisted ISAC systems by exploiting the sensing ability of the BS. The contributions of this paper are mainly summarized as follows.

\begin{itemize}
%\item We find one key difference between the conventional communication system and the ISAC system lies in the full-duplex mode and sensing ability of the ISAC BS, which allows the ISAC BS to directly perform beam training between BS and RIS and obtain the position of the RIS based on the echoes from the RIS. 

%However, different from the work in~\cite{icc23CKJ} that is only adapted to the line-of-sight (LoS) channels, the SBTTS scheme in this paper integrating the new proposed iterative parameter estimation for beam training and target sensing (IPEBTTS) algorithm is adapted to both the LoS channels and non-LoS (NLoS) channels.

\item We first propose a simultaneous beam training and target sensing (SBTTS) scheme, which enables the BS to perform beam training and simultaneously to sense the targets. Based on our findings, the energy of the echoes from the RIS is accumulated in the angle-delay domain while that from the targets is accumulated in the Doppler-delay domain. The SBTTS scheme can distinguish the RIS from the targets with the mixed echoes from the RIS and the targets.

\item We propose a positioning and array orientation estimation (PAOE) scheme for both the line-of-sight (LoS) channels and the non-line-of-sight (NLoS) channels based on the beam training results of SBTTS, where a low-complexity two-dimensional fast search (TDFS) algorithm is proposed to obtain the position and array orientation for the NLoS channels.

\item Based on the SBTTS and PAOE schemes, we further compute the angle-of-arrival (AoA) and angle-of-departure (AoD) for the channels between the RIS and the UTs by exploiting the geometry relationship to accomplish the beam alignment of the RIS-assisted ISAC systems, substantially reducing the training overhead.

\end{itemize}

The rest of this paper is organized as follows. Section~\ref{SystemModel} introduces the system model of the RIS-assisted ISAC systems. In Section~\ref{SBT}, we propose the SBTTS scheme. The PAOE  scheme is provided in Section~\ref{PAO}. Beam alignment based on SBTTS and PAOE schemes is discussed in Section~\ref{SBA}. The simulation results are presented in Section~\ref{SimulationResults}, and the paper is concluded in Section~\ref{Conclusion}.

The notations are defined as follows. Symbols for matrices (upper case) and vectors (lower case) are in boldface. $(\cdot)^{\rm T}$ and $(\cdot)^{\rm H} $ denote the matrix or vector transpose and conjugate transpose (Hermitian), respectively. $[\boldsymbol{a}]_{n}$, $\left[ \boldsymbol{A} \right] _{:,n}$ and $\left[ \boldsymbol{A} \right] _{m,n}$ denote the $n$th entry of vector $\boldsymbol{a}$, the $n$th column of matrix $\boldsymbol{A}$ and the entry on the $m$th row and the $n$th column of matrix $\boldsymbol{A}$, respectively. ${\rm diag}\{\boldsymbol{A}\}$ denotes the vector consisting of the diagonal entries of matrix $\boldsymbol{A}$ and ${\rm diag}\{\boldsymbol{a}\}$ denotes the diagonalization operation of vector $\boldsymbol{a}$. $\mathbb{N}$, $\mathbb{C}$, $\mathcal{C}\mathcal{N}$,  $U$, $\boldsymbol{I}_{N}$, $\mathcal{O}$, and $j$ denote the set of integer, the set of complex number, the complex Gaussian distribution, the uniform random distribution, an $N\times N$ identity matrix, the order of complexity, and the square root of $-1$, respectively. ${\rm sgn}(\cdot)$, ${\rm asin(\cdot)}$, and $\mathrm{vec}(\cdot)$ denote the sign function, the arcsine function, and the vectorization operation, respectively. In addition, $|\cdot |$ and $\|\cdot \|_2$ denote the absolute value of a scalar and $\ell_2$-norm of a vector, respectively. 

\begin{figure}[!t]
	\centering
	\includegraphics[width=70mm]{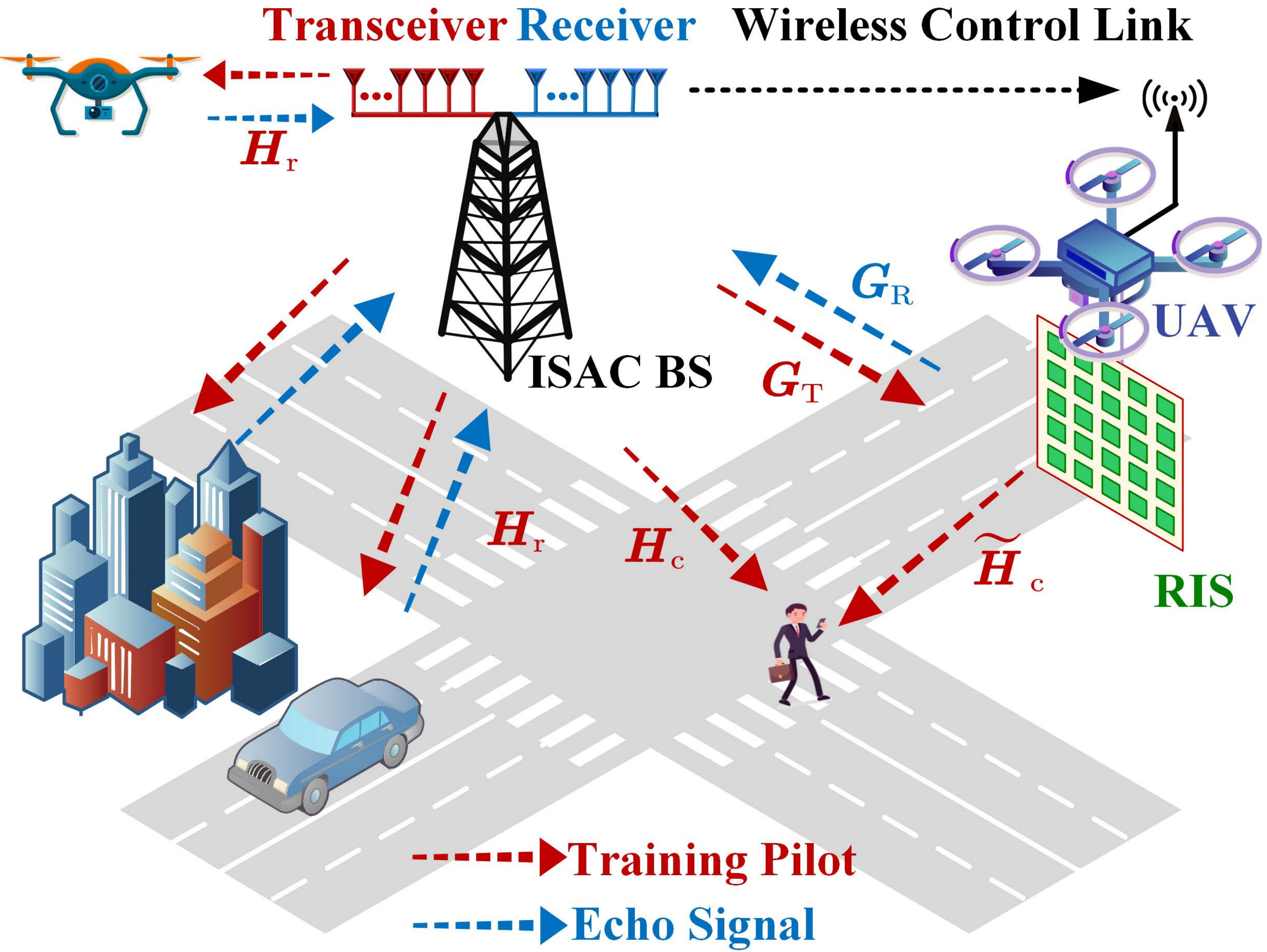}
	\caption{Illustration of the  beam training and target sensing in the ISAC.}
	\label{FigSystemModel}
	\vspace{-0.8cm}
\end{figure}

\section{System Model}\label{SystemModel}
As shown in Fig.~\ref{FigSystemModel}, we consider a RIS-assisted ISAC system, where an ISAC BS serves a UT and detects several targets simultaneously, under the assistance of a RIS. The BS includes a communication transceiver equipped with $N_{\rm T}$ antennas and an echo signal receiver equipped with $N_{\rm R}$ antennas, where the antennas are placed in uniform linear arrays (ULAs) with half wavelength intervals. To reduce the hardware costs, the fully-connected hybrid beamforming is adopted at both the communication transceiver and the echo receiver. In this work, we consider the monostatic ISAC scenario and assume the transceiver as well as the receiver to be collocated. At the RIS, $N_{\rm RIS}$ RIS elements are deployed for passive reflection and a RIS controller with wireless control link to the BS is installed, allowing the BS to control the RIS in real time. For the UT, a  half-wavelength-interval ULA with $N_{\rm UT}$ antennas is equipped and the analog combining is adopted to form directional beams. 

We assume orthogonal frequency division multiplexing (OFDM) with $M$ subcarriers is employed for wideband processing. During the beam training and target sensing, the transceiver transmits $P$ OFDM training symbols continuously. Meanwhile, the UT receives the downlink training symbols  and the collocated receiver collects the echoes from the RIS as well as the targets. To combat the multipath effects and eliminate the inter-symbol interference, the cyclic prefix (CP) is employed for each OFDM symbol and can be easily removed at the receiver by exploiting the orthogonality of the subcarriers~\cite{JSAC22Qinshi}. Then, for the $m$th subcarrier of the $p$th OFDM symbol, the received signal without the CP by the BS can be expressed as 
\begin{equation}\label{SystemModelofBS}
\setlength{\abovedisplayskip}{3pt}
\setlength{\belowdisplayskip}{3pt}
{y}_{m}^{(p)} = \boldsymbol{v}_{p}^{\rm H} \left({\boldsymbol{G}_{{\rm R},m}}\boldsymbol{\Phi}_{p}\boldsymbol{G}_{{\rm T},m}\!+\!\boldsymbol{H}_{{\rm r},m}^{(p)}\right)\boldsymbol{f}_{p} x_{m}^{(p)} + \boldsymbol{v}_{p}^{\rm H} \boldsymbol{\eta}_{{\rm r},m}^{(p)},
\end{equation}
where $\boldsymbol{v}_{p}\in \mathbb{C}^{N_{\rm R}\times 1}$ denotes the analog beamformer at the receiver, $\boldsymbol{G}_{{\rm R},m}\in\mathbb{C}^{ N_{\rm R}\times N_{\rm RIS}}$ denotes the communication  channel between the RIS and the receiver, $\boldsymbol{\Phi}_{p} \triangleq {\rm diag}\{[e^{j\phi_1},e^{j\phi_2},\cdots,e^{j\phi_{N_{\rm RIS}}}]^{\rm T}\}$ denotes the reflection coefficients of the RIS elements, $\boldsymbol{G}_{{\rm T},m}\in\mathbb{C}^{N_{\rm RIS} \times N_{\rm T}}$ denotes the communication channel between the transceiver and the RIS, $\boldsymbol{H}_{{\rm r},m}^{(p)}$ denotes the sensing channel of the $T$ targets, $\boldsymbol{f}_{p}\in\mathbb{C}^{N_{\rm T}\times 1}$ denotes the analog beamformer at the transceiver, $x_{m}^{(p)}$ denotes the transmit signal, and $ \boldsymbol{\eta}_{{\rm r},m}^{(p)}\sim\mathcal{C}\mathcal{N}\left( \boldsymbol{0},\sigma_{\rm r}^{2}\boldsymbol{I}_{N_{\rm R}} \right)$
denotes the additive white Gaussian noise (AWGN) vector. 

\textbf{Remark 1:} In practice,  the signal received by the BS also contains the clutters. To avoid the influence of these clutters, a common approach is to estimate the clutters in prior and remove them from the received signal via background subtraction~\cite{JSAC22SXD,TAES22EML} or filtering~\cite{TGRS09YY,FRSP}. Therefore, in this work, we assume the clutters have been removed and focus on the signal of interest.

%the reflections of clutters are different from the noise in that the former is identical for the same subcarrier during the coherent time while the later varies for %different signal receptions

On a parallel branch,  for the $m$th subcarrier of the $p$th symbol, the received signal at the UT can be expressed as
\begin{equation}\label{SystemModelofUT}
\setlength{\abovedisplayskip}{3pt}
\setlength{\belowdisplayskip}{3pt}
	{z}_{m}^{(p)} = \boldsymbol{w}_{p}^{\rm H} (\boldsymbol{H}_{{\rm c},m} + \widetilde{\boldsymbol{H}}_{{\rm c},m}\boldsymbol{\Phi}_p\boldsymbol{G}_{{\rm T},m}) \boldsymbol{f}_{p} x_{m}^{(p)} + \boldsymbol{w}_{p}^{\rm H}\boldsymbol{\eta}_{{\rm c},m}^{(p)},
\end{equation}
where $\boldsymbol{w}_{p}\in \mathbb{C}^{N_{\rm UT}\times 1}$ denotes the analog beamformer, $\boldsymbol{H}_{{\rm c},m}\in \mathbb{C}^{N_{\rm UT} \times N_{\rm T}}$ denotes the channel between the transceiver and the UT, $\widetilde{\boldsymbol{H}}_{{\rm c},m} \in \mathbb{C}^{N_{\rm UT} \times N_{\rm RIS}}$ denotes the channel between the RIS and the UT, and $\boldsymbol{\eta}_{{\rm c},m}^{(p)}\sim\mathcal{C}\mathcal{N}\left( \boldsymbol{0},\sigma_{\rm c}^{2}\boldsymbol{I}_{N_{\rm UT}} \right)$ denotes the AWGN vector. Note that the AWGN vectors, $\boldsymbol{\eta}_{{\rm r},m}^{(p)}$ and $\boldsymbol{\eta}_{{\rm c},m}^{(p)}$, are independent across both OFDM subcarriers and symbols.

\textbf{Remark 2:} In \eqref{SystemModelofBS} and \eqref{SystemModelofUT}, we omit the digital beamformers at both the communication transceiver and the echo signal receiver because we only use one RF chain that is enough to acquire the CSI of the RIS-assisted ISAC systems. However, multiple RF chains can also be exploited to support parallel transmission of multiple data streams after the CSI is acquired.

For the wideband OFDM systems, the frequency-domain communication channel between the transceiver and the RIS at the $m$th subcarrier can be expressed as~\cite{JSTSP21ZhangAndrew}
\begin{equation}\label{ChannelmodelofRIS}
\setlength{\abovedisplayskip}{3pt}
\setlength{\belowdisplayskip}{3pt}
	\boldsymbol{G}_{{\rm T},m} = \sum_{l=1}^{L_{\rm BR}} \zeta_{m}^{(l)}\boldsymbol{\alpha}\left(N_{\rm RIS},\varphi_{\rm BR}^{(l)}\right)\boldsymbol{\alpha}\left(N_{\rm T},\theta_{\rm BR}^{(l)}\right)^{\rm H},
\end{equation}
where $\zeta_{m}^{(l)}\triangleq g_{\rm BR}^{(l)} e^{-j2\pi (m-1) \tau_{\rm BR}^{(l)}\Delta f}$;  $g_{\rm BR}^{(l)}$, $\tau_{\rm BR}^{(l)}$, $\varphi_{\rm BR}^{(l)}$ and $\theta_{\rm BR}^{(l)}$ denote the channel gain, the channel delay, the channel AoA and  the channel AoD of the $l$th path, respectively. $L_{\rm BR}$ denotes the number of paths, and $\Delta f$ denotes the subcarrier spacing. $\tau_{\rm BR}^{(l)} = r_{\rm BR}^{(l)}/c$, where  $r_{\rm BR}^{(l)}$ denotes the distance of the $l$th path from the transceiver to the RIS and $c$ denotes the speed of light.  $\boldsymbol{\alpha}(\cdot)$ denotes the   steering vector and can be expressed as
%Define the physical channel AoD and AoA of the $l$th path as $\omega_{\rm BR}^{(l)}$ and $\vartheta_{\rm BR}^{(l)}$, respectively, where $\omega_{\rm BR}^{(l)} \in [-90^{\circ},90^{\circ}]$ and $\vartheta_{\rm BR}^{(l)} \in [-90^{\circ},90^{\circ}]$. We have $\theta_{\rm BR}^{(l)} = \sin(\omega_{\rm BR}^{(l)})$ and $\varphi_{\rm BR}^{(l)} = \sin(\vartheta_{\rm BR}^{(l)})$ considering the half wavelength interval of antennas. Then, we have $\theta_{\rm BR}^{(l)}\in [-1,1]$ and $\varphi_{\rm BR}^{(l)}\in [-1,1]$.
\begin{equation}\label{Channelsteeringvector}
\setlength{\abovedisplayskip}{3pt}
\setlength{\belowdisplayskip}{3pt}
 	\boldsymbol{\alpha}(N,\theta) = \frac{1}{\sqrt{N}}[1,e^{j\pi\theta},\cdots e^{j(N-1)\pi\theta}]^{\rm T}.
\end{equation}
According to the channel reciprocity, we have
\begin{equation}\label{ChannelmodelofRIS2}
	\boldsymbol{G}_{{\rm R},m} = \sum_{l=1}^{L_{\rm BR}} \zeta_{m}^{(l)} \boldsymbol{\alpha}\left(N_{\rm R},\theta_{\rm BR}^{(l)}\right)\boldsymbol{\alpha}\left(N_{\rm RIS},\varphi_{\rm BR}^{(l)}\right)^{\rm T}.
\end{equation}
Similarly, for the wideband OFDM systems, the frequency-domain communication channel between the transceiver and the UT at the $m$th subcarrier can be expressed as
\begin{equation}\label{ChannelmodelofUT1}
	\boldsymbol{H}_{{\rm c},m} = \sum_{l=1}^{L_{\rm BU}} \chi_{m}^{(l)} \boldsymbol{\alpha}\left(N_{\rm UT},\varphi_{\rm BU}^{(l)}\right)\boldsymbol{\alpha}\left(N_{\rm T},\theta_{\rm BU}^{(l)}\right)^{\rm H},
\end{equation}
where $\chi_{m}^{(l)} \triangleq g_{\rm BU}^{(l)} e^{-j2\pi (m-1) \tau_{\rm BU}^{(l)}\Delta f}$; $g_{\rm BU}^{(l)}$, $\tau_{\rm BU}^{(l)}$, $\varphi_{\rm BU}^{(l)}$ and $\theta_{\rm BU}^{(l)}$ denote the channel gain, the channel delay, the channel AoA, and the channel AoD of the $l$th path, respectively. $L_{\rm BU}$ denotes the number of paths. $\tau_{\rm BU}^{(l)} = r_{\rm BU}^{(l)}/c$, where  $r_{\rm BU}^{(l)}$ denotes the distance of the $l$th path from the transceiver to the UT. Moreover, the frequency-domain communication channel between the RIS and the UT at the $m$th subcarrier can be expressed as
\begin{equation}\label{ChannelmodelofUT2}
	\widetilde{\boldsymbol{H}}_{{\rm c},m} = \sum_{l=1}^{L_{\rm RU}} \widetilde{\chi}_{m}^{(l)} \boldsymbol{\alpha}\left(N_{\rm UT},\varphi_{\rm RU}^{(l)}\right)\boldsymbol{\alpha}\left(N_{\rm RIS},\theta_{\rm RU}^{(l)}\right)^{\rm T},
\end{equation}
where $\widetilde{\chi}_{m}^{(l)} \triangleq g_{\rm RU}^{(l)} e^{-j2\pi (m-1) \tau_{\rm RU}^{(l)}\Delta f}$; $g_{\rm RU}^{(l)}$, $\tau_{\rm RU}^{(l)}$, $\varphi_{\rm RU}^{(l)}$ and $\theta_{\rm RU}^{(l)}$ denote the channel gain, the channel delay, the channel AoA and the channel AoD of the $l$th path, respectively. $L_{\rm RU}$ denotes the number of paths. $\tau_{\rm RU}^{(l)} = r_{\rm RU}^{(l)}/c$, where  $r_{\rm RU}^{(l)}$ denotes the distance of the $l$th path from the RIS to the UT.

The sensing channel of targets can be expressed as
\begin{equation}\label{ChannelModelofTargets}
	\boldsymbol{H}_{{\rm r},m}^{(p)} = \sum_{l=1}^{T} \gamma_{m,p}^{(l)} \boldsymbol{\alpha}\left(N_{\rm R},\theta_{\rm Tar}^{(l)}\right)\boldsymbol{\alpha}\left(N_{\rm T},\theta_{\rm Tar}^{(l)}\right)^{\rm H},
\end{equation}
where $\gamma_{m,p}^{(l)} \triangleq g_{\rm Tar}^{(l)} e^{-j2\pi (m-1) \tau_{\rm{Tar}}^{(l)}\Delta f}e^{j2\pi  (p-1) f_{\rm Tar}^{(l)} T_{\rm s}}$; $T$,  $g_{\rm Tar}^{(l)}$, $\tau_{\rm{Tar}}^{(l)}$, $f_{\rm Tar}^{(l)}$ and $\theta_{\rm Tar}^{(l)}$ denote the number of targets, the channel gain, the channel delay, the Doppler frequency and the channel AoD of the $l$th target, respectively. $T_{\rm s}$ denotes the duration of the OFDM symbol. $\tau_{\rm{Tar}}^{(l)} = 2r_{\rm Tar}^{(l)}/c$, where $r_{\rm Tar}^{(l)}$ denotes the distance from the transceiver to the $l$th target. $f_{\rm Tar}^{(l)} = 2 v_{\rm Tar}^{(l)}/\lambda$, where $v_{\rm Tar}^{(l)}$ denotes the radial velocity of the $l$th target and $\lambda$ denotes the wavelength of the carrier frequency.
% In the existing literature, the two parts of beam training are performed separately by switching the working status of the RIS~\cite{TWC20WZR}. %Since the beam training for the BS-UT link has been extensively investigated in the exsiting literature, we focus on the beam training for the link of BS-RIS-UT.

To support high-speed transmission, beam alignment among the BS, the RIS, and the UT is needed. One way to achieve high-quality beam alignment is the codebook-based beam training~\cite{WC22WPL}. Generally, two parts of beam training, including the beam training for the BS-UT link and the beam training for the BS-RIS-UT link, are needed  to accomplish beam alignment among the BS, the RIS, and the UT. To reveal the inherent shortcomings of the conventional beam training method, we take the beam training for the BS-RIS-UT link as an example. The codebooks for the BS, the RIS, and the UT are expressed as $\boldsymbol{\mathcal{B}} = \{\boldsymbol{b}_1,\boldsymbol{b}_2,\cdots,\boldsymbol{b}_{N_{\rm T}}\}$,  $\boldsymbol{\mathcal{R}} = \{\boldsymbol{r}_1,\boldsymbol{r}_2,\cdots,\boldsymbol{r}_{N_{\rm RIS}}\}$, and $\boldsymbol{\mathcal{U}} = \{\boldsymbol{u}_1,\boldsymbol{u}_2,\cdots,\boldsymbol{u}_{N_{\rm UT}}\}$, respectively, where
\begin{align}\label{codebook}
	&\boldsymbol{b}_n = \boldsymbol{\alpha}(N_{\rm T},(2n-2)/N_{\rm T}),~n=1,2,\cdots,N_{\rm T}, \nonumber\\
	&\boldsymbol{r}_s = \sqrt{N_{\rm RIS}}\boldsymbol{\alpha}(N_{\rm RIS},(2s-2)/N_{\rm RIS}),~s=1,2,\cdots,N_{\rm RIS},\nonumber\\
	&\boldsymbol{u}_t = \boldsymbol{\alpha}(N_{\rm UT},(2t-2)/N_{\rm UT}),~t=1,2,\cdots,N_{\rm UT}.
\end{align}
The existing works focus on \eqref{SystemModelofUT} and perform the beam alignment for the BS-RIS-UT link via
\begin{align}\label{RISbeamtraining}
	&\max_{\boldsymbol{w}_{p},\boldsymbol{\Phi}_p,\boldsymbol{f}_{p},\varpi} \left|\sum_{m=1}^{M}e^{j(m-1)\varpi}\boldsymbol{w}_{p}^{\rm H}\widetilde{\boldsymbol{H}}_{{\rm c},m}\boldsymbol{\Phi}_p\boldsymbol{G}_{{\rm T},m}\boldsymbol{f}_{p}\right|^2\nonumber\\
	&{\rm~~~~~s.t.}~~~~\boldsymbol{w}_{p}\in\boldsymbol{\mathcal{U}},~{\rm diag}\{\boldsymbol{\Phi}_p\}\in\boldsymbol{\mathcal{R}},~\boldsymbol{f}_{p}\in\boldsymbol{\mathcal{B}}, \varpi \in [0,2\pi],
\end{align}
where $e^{j(m-1)\varpi}$ is the introduced phase factor to accumulate the energy of all subcarriers and can be designed by performing the Fourier transform of the signals at $M$ subcarriers. To solve \eqref{RISbeamtraining}, we need to test all the codewords in $\boldsymbol{\mathcal{B}}$, $\boldsymbol{\mathcal{R}}$ and $\boldsymbol{\mathcal{U}}$ one by one. Totally $N_{\rm T}N_{\rm RIS}N_{\rm UT}$ times of beam training are needed, which is much larger than the $N_{\rm T}N_{\rm UT}$ times of beam training for the BS-UT link. Novel beam training scheme for the BS-RIS-UT link with low training overhead is desired. On the other hand, in the emerging ISAC systems, the BS is endowed with the ability of sensing. It would be interesting to investigate how to exploit the  sensing ability of the BS for reducing the training overhead of the BS-RIS-UT link, which will be discussed in the next sections.

% Our objective for beam training is to solve \eqref{RISbeamtraining} with considerable reduction in training overhead.

During the beam training, beam alignment needs to be accomplished at both the transmitter and the receiver. As a result, the transmitter needs to transmit multiple pilots with the same beamforming vector so that the receiver can test the codewords in its codebook sequentially. On the other hand, during the target sensing, the transmitter also needs to transmit continuous  waveform with the same beamforming vector so that the Doppler effects can be exploited to identify the moving targets. The similarity between the beam training and the target sensing implies that we may reuse the downlink beam training pilots for target sensing, which will also be discussed in the next sections.

\section{Simultaneous Beam Training and Target Sensing Scheme} \label{SBT}

In this section, we propose the SBTTS scheme. Based on our findings, the energy of the echoes from the RIS is accumulated in the angle-delay domain while that from the targets is accumulated in the Doppler-delay domain.  The SBTTS scheme  integrating the IPEBTTS algorithm  can distinguish the RIS from the targets with the mixed echoes from the RIS and the targets.

% integrating the new proposed iterative parameter estimation for beam training and target sensing (IPEBTTS) algorithm

\subsection{Angle-Delay Domain vs Doppler-Delay Domain }\label{SecIIIA}
%We first focus on the beam alignment for the BS-RIS-UT link,  which includes the BS-RIS link and the RIS-UT link. 
Note that the existing works perform beam alignment for the BS-RIS-UT link via solving \eqref{RISbeamtraining}. Different from the existing works, we first turn to the BS-RIS-BS link contained in \eqref{SystemModelofBS} and perform the beam alignment between the BS and the RIS via
\begin{align}\label{BSRISbeamtraining}
	&\max_{~\boldsymbol{\Phi}_p,\boldsymbol{f}_{p},\varpi} \left|\sum_{m=1}^{M}e^{j(m-1)\varpi}\boldsymbol{v}_{p}^{\rm H}\boldsymbol{G}_{{\rm R},m}\boldsymbol{\Phi}_p\boldsymbol{G}_{{\rm T},m}\boldsymbol{f}_{p}\right|^2\nonumber\\
&{\rm ~~~s.t.}~~~~{\rm diag}\{\boldsymbol{\Phi}_p\}\in\boldsymbol{\mathcal{R}},~\boldsymbol{f}_{p}\in\boldsymbol{\mathcal{B}},\nonumber \\
& ~~~~~~~~~~~\boldsymbol{v}_{p} = \boldsymbol{\alpha}(N_{\rm R},\varXi(\boldsymbol{f}_{p})), \varpi \in [0,2\pi],
\end{align}
where $\varXi(\cdot)$ denotes the spatial direction of a steering vector, that is, $\varXi(\boldsymbol{\alpha}(N,\theta)) = \theta$. To solve \eqref{BSRISbeamtraining}, we test all combinations of the codewords in $\boldsymbol{\mathcal{B}}$ and $\boldsymbol{\mathcal{R}}$, and only $N_{\rm T}N_{\rm RIS}$ times of beam training are needed thanks to the channel reciprocity between the BS-RIS link  and the RIS-BS link. When testing the $n$th codeword in $\boldsymbol{\mathcal{B}}$ and the $s$th codeword in $\boldsymbol{\mathcal{R}}$, the received signal by the BS can be expressed as
\begin{equation}\label{BeamTrainingModel}
	{y}_{m}^{(p)} = \boldsymbol{v}_{n}^{\rm H} ({\boldsymbol{G}_{{\rm R},m}}{\rm diag}\{\boldsymbol{r}_s\}\boldsymbol{G}_{{\rm T},m}\!+\!\boldsymbol{H}_{{\rm r},m}^{(p)})\boldsymbol{b}_{n} + \boldsymbol{v}_{n}^{\rm H}  \boldsymbol{\eta}_{{\rm r},m}^{(p)},
\end{equation}
where $n \in \{1,2,\cdots N_{\rm T}\} $, $s\in \{1,2,\cdots,N_{\rm RIS}\} $, $\boldsymbol{v}_{n} = \boldsymbol{\alpha}(N_{\rm R},\varXi(\boldsymbol{b}_{n}))$, and $p = (n-1)N_{\rm RIS} + s$. In \eqref{BeamTrainingModel}, we set $x_{m}^{(p)} = 1$. Note that $N_{\rm RIS}$ codewords in  $\boldsymbol{\mathcal{R}}$ are tested sequentially when testing each codeword in $\boldsymbol{\mathcal{B}}$. To test the $n$th codeword in $\boldsymbol{\mathcal{B}}$, we stack the $N_{\rm RIS}$ times of receptions at the $M$ subcarriers together as $\boldsymbol{Y}_n \in \mathbb{C}^{N_{\rm RIS}\times M}$. Then, we have
\begin{align}\label{BeamTrainingStack}
	\boldsymbol{Y}_n &= {{\sum_{r=1}^{L_{\rm BR}} \sum_{u=1}^{L_{\rm BR}} \mu_{r,u} \boldsymbol{F}_{N_{\rm RIS}} \boldsymbol{\alpha}\left(N_{\rm RIS},\varphi_{\rm BR}^{(r)}+\varphi_{\rm BR}^{(u)}\right)\boldsymbol{e}_{r,u}^{\rm T}}}+ \boldsymbol{N}_n  \nonumber\\
	&~~~+\sum_{l=1}^{T} \kappa_l \boldsymbol{\alpha}\left(N_{\rm RIS},2f_{\rm Tar}^{(l)}T_{\rm s}\right)\boldsymbol{\alpha}\left(M,-2\tau_{\rm Tar}^{(l)}\Delta f\right)^{\rm T} \nonumber\\
	&\overset{\rm (a)}{\approx} {{\sum_{u=1}^{L_{\rm BR}} {\mu}_{u,u} \boldsymbol{F}_{N_{\rm RIS}} \boldsymbol{\alpha}\left(N_{\rm RIS},2\varphi_{\rm BR}^{(u)}\right)\boldsymbol{e}_{u,u}^{\rm T}}}+ \boldsymbol{N}_n  \nonumber\\
	&~~~+\sum_{l=1}^{T}\!\kappa_l\boldsymbol{\alpha}\!\left(\!N_{\rm RIS},2f_{\rm Tar}^{(l)}T_{\rm s}\!\right)\!\boldsymbol{\alpha}\left(\!M,-2\tau_{\rm Tar}^{(l)}\Delta f\!\right)^{\rm T},
\end{align}
where
\begin{align}\label{BeamTrainingStackVar}
	&\mu_{r,u} = g_{\rm BR}^{(r)} g_{\rm BR}^{(u)} \boldsymbol{v}_n^{\rm H} \boldsymbol{\alpha}\left(N_{\rm R},\theta_{\rm BR}^{(r)}\right) \boldsymbol{\alpha}\left(N_{\rm T},\theta_{\rm BR}^{(u)}\right)^{\rm H}\boldsymbol{b}_n, \nonumber\\
	&\kappa_l = \sqrt{MN_{\rm RIS}}g_{\rm Tar}^{(l)}\boldsymbol{v}_n^{\rm H} \boldsymbol{\alpha}\left(N_{\rm R},\theta_{\rm Tar}^{(l)}\right) \boldsymbol{\alpha}\left(N_{\rm T},\theta_{\rm Tar}^{(l)}\right)^{\rm H}\boldsymbol{b}_n,\nonumber\\
	&\boldsymbol{F}_{N_{\rm RIS}} = [\boldsymbol{r}_1,\boldsymbol{r}_2,\cdots,\boldsymbol{r}_{N_{\rm RIS}}]^{\rm T},\nonumber\\
	&\boldsymbol{e}_{r,u} = \sqrt{M}\boldsymbol{\alpha}\left(M,-2\left(\tau_{\rm BR}^{(r)} + \tau_{\rm BR}^{(u)}\right)\Delta f\right),
\end{align}
and $\boldsymbol{N}_n$ is the stack of noise terms. In \eqref{BeamTrainingStack}, (a) follows $\mu_{r,u} \approx 0$ if $\big|\theta_{\rm BR}^{(u)} - (2n-2)/N_{\rm T}\big| > 1/N_{\rm T}$ or $\big|\theta_{\rm BR}^{(r)} - (2n-2)/N_{\rm T}\big| > 1/N_{\rm R}$.

Note that both $\boldsymbol{e}_{u,u}$ and $\boldsymbol{\alpha}\left(M,-2\tau_{\rm Tar}^{(l)}\Delta f\right)$ are the steering vectors related to both the subcarrier spacing and the channel delay. A regular approach to accumulate  the energy of the received signals at all subcarriers is the Fourier transform. Denote the $M\times M$ inverse discrete Fourier transform (IDFT) matrix as $\boldsymbol{F}_M$, where $[\boldsymbol{F}_M]_{:,m} = \sqrt{M}\boldsymbol{\alpha}(M,(2m-2)/M)$. Postmutiplying $\boldsymbol{Y}_n$ with $\boldsymbol{F}_M$, we can obtain
\begin{align}\label{BeamTraining1DFT}
	\widetilde{\boldsymbol{Y}}_n &= {{\sum_{u=1}^{L_{\rm BR}} \mu_{u,u} \boldsymbol{F}_{N_{\rm RIS}} \boldsymbol{\alpha}\left(N_{\rm RIS},2\varphi_{\rm BR}^{(u)}\right)\widetilde{\boldsymbol{e}}_{u,u}^{\rm T}}} \nonumber\\
	&~~~+ \sum_{l=1}^{T} \kappa_l \boldsymbol{\alpha}\left(N_{\rm RIS},2f_{\rm Tar}^{(l)}T_{\rm s}\right)\widetilde{\boldsymbol{d}}_{l}^{\rm T}   + \widetilde{\boldsymbol{N}}_n,
\end{align}
where
\begin{align}\label{BeamTraining1DFTVar}
	&[\widetilde{\boldsymbol{e}}_{u,u}]_m = G_M\left({(2m-2)\pi}/{M}-4\pi\tau_{\rm BR}^{(u)}\Delta f\right), \nonumber\\
	&[\widetilde{\boldsymbol{d}}_{l}]_m = G_M\left({(2m-2)\pi}/{M}-2\pi \tau_{\rm Tar}^{(l)} \Delta f\right)/\sqrt{M}, \nonumber\\
	&G_M(\psi) \triangleq e^{j(M-1)\psi/2}\sin(M\psi/2)/\sin(\psi/2),
\end{align}
and $\widetilde{\boldsymbol{N}}_n = {\boldsymbol{N}}_n \boldsymbol{F}_M$. In \eqref{BeamTraining1DFT}, the peaks of RIS echoes from different angles locate at different rows of  $\widetilde{\boldsymbol{Y}}_n$. In addition, the peaks of echoes from either RIS or targets with different delays locate at different columns of $\widetilde{\boldsymbol{Y}}_n$ regarding the property of $G_M(\psi)$. Therefore, \eqref{BeamTraining1DFT} is an expression in the angle-delay domain.

\begin{figure*}[htbp]
	\centering
	\subfigure[Illustration of the signals in the angle-delay domain.]
	{
		\begin{minipage}[b]{.48\linewidth}\label{FigAD}
			\centering
			\includegraphics[scale=0.48]{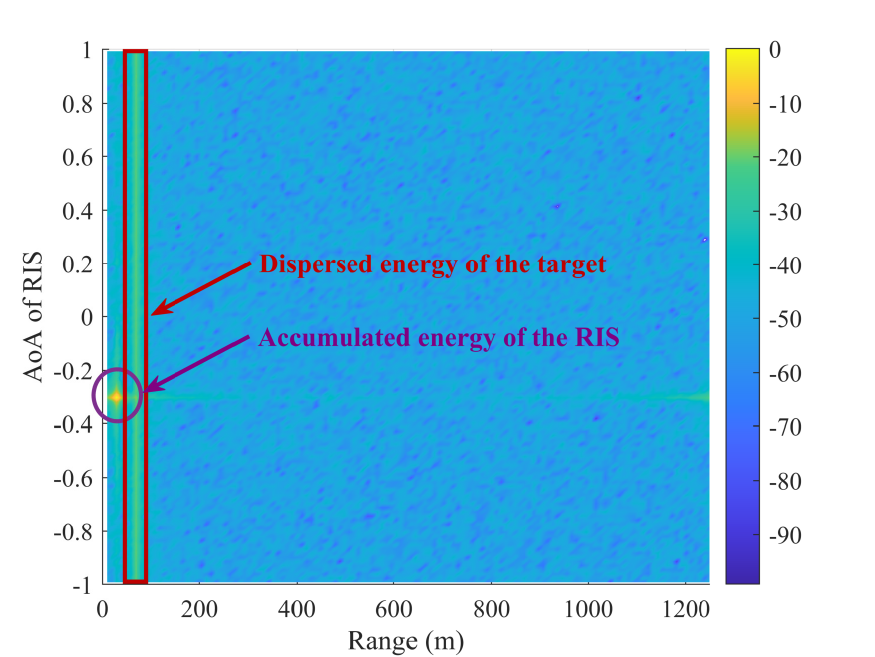}
		\end{minipage}
	}
	\subfigure[Illustration of the signals in the Doppler-delay domain.]
	{
		\begin{minipage}[b]{.48\linewidth}\label{FigDD}
			\centering
			\includegraphics[scale=0.48]{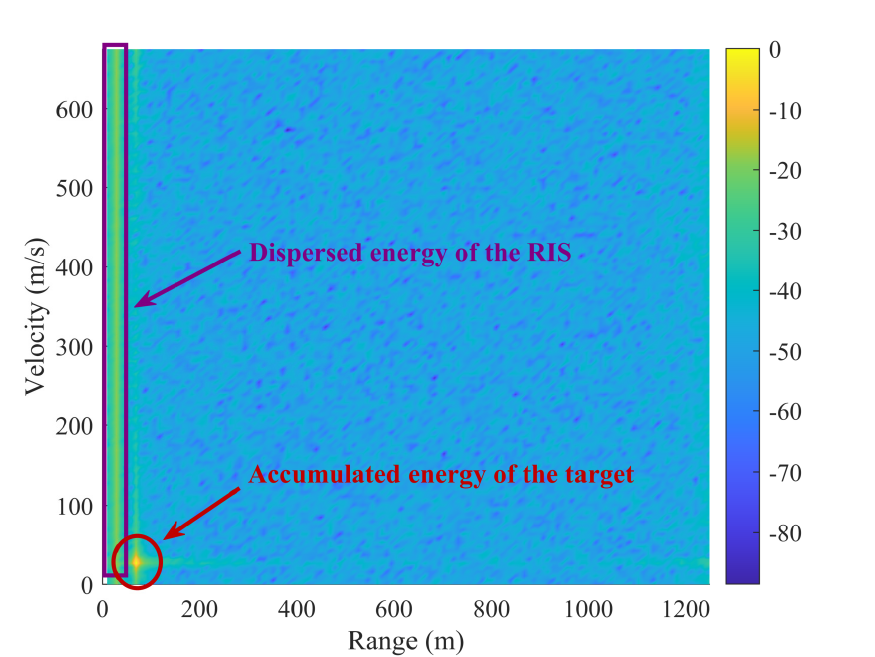}
		\end{minipage}
	}
	\caption{Illustration of signals in the angle-delay domain and the Doppler-delay domain. We set $N_{\rm T} = 32$, $N_{\rm R}  = 32$, $N_{\rm RIS} = 128$, $L_{\rm BR} = 1$, $T = 1$, $M = 128$, $\Delta f = 120$ kHz. The transmit power and the noise power are set to be $50$~dBm and $-103$~dBm, respectively.}
	\label{FigADDD}
\end{figure*}

Note that $\boldsymbol{\alpha}\left(N_{\rm RIS},2f_{\rm Tar}^{(l)}T_{\rm s}\right)$ in \eqref{BeamTraining1DFT} is a steering vector related to the Doppler frequency and the OFDM symbol duration. To accumulate the energy of the echoes from the targets over time, we denote the $N_{\rm RIS}\times N_{\rm RIS}$ DFT matrix as $\boldsymbol{W}_{N_{\rm RIS}}$, where $[\boldsymbol{W}_{N_{\rm RIS}}]_{:,n} = \sqrt{{N_{\rm RIS}}}\boldsymbol{\alpha}({N_{\rm RIS}},(2-2n)/{N_{\rm RIS}})$. Premutiplying $\widetilde{\boldsymbol{Y}}_n$ with $\boldsymbol{W}_{N_{\rm RIS}}^{\rm T}/\sqrt{N_{\rm RIS}}$, we have 
\begin{align}\label{BeamTraining2DFT}
	\overline{\boldsymbol{Y}}_n &\overset{\rm (a)}{=} { \sum_{u=1}^{L_{\rm BR}} \mu_{u,u}  \boldsymbol{\alpha}\left(N_{\rm RIS},2\varphi_{\rm BR}^{(u)}\right)\widetilde{\boldsymbol{e}}_{u,u}^{\rm T}} + \sum_{l=1}^{T} \kappa_l \widetilde{\boldsymbol{c}}_l\widetilde{\boldsymbol{d}}_{l}^{\rm T}   + \overline{\boldsymbol{N}}_n,
\end{align}
where
\begin{align}\label{BeamTraining2DFTS} 
	[\widetilde{\boldsymbol{c}}_l]_s = \frac{1}{N_{\rm RIS}}G_{N_{\rm RIS}}\left(2\pi f_{\rm Tar}^{(l)}T_{\rm s} - \frac{(2s-2)\pi}{N_{\rm RIS}} \right),
\end{align}
$\overline{\boldsymbol{N}}_n = \boldsymbol{W}_{N_{\rm RIS}}^{\rm T}\widetilde{\boldsymbol{N}}_n/\sqrt{N_{\rm RIS}}$, and (a) follows $\boldsymbol{W}_{N_{\rm RIS}}^{\rm T}\boldsymbol{F}_{N_{\rm RIS}}/\sqrt{N_{\rm RIS}} = \boldsymbol{I}_{N_{\rm RIS}}$. In \eqref{BeamTraining2DFT}, the physical meaning  of the columns of $\overline{\boldsymbol{Y}}_n$  is the same as that of $\widetilde{\boldsymbol{Y}}_n$ in \eqref{BeamTraining1DFT}. However, the physical meaning  of the rows is different. To be specific, the peaks of echoes from targets with different Doppler frequencies locate at different rows of  $\overline{\boldsymbol{Y}}_n$ regarding the property of $G_M(\psi)$. Therefore, \eqref{BeamTraining2DFT} is an expression in the Doppler-delay domain. 

%\begin{figure}[!t]
%	\centering  %图片全局居中
%	\subfigure[Illustration of the signals in the angle-delay domain.]{
%		\label{FigAD}
%		\includegraphics[width=0.45\textwidth]{ADDomainDulp.pdf}}
%	\subfigure[Illustration of the signals in the Doppler-delay domain.]{
%		\label{FigDD}
%		\includegraphics[width=0.45\textwidth]{DDDomainDulp.pdf}}
%	\caption{Illustration of signals in the angle-delay domain and the Doppler-delay domain. We set $N_{\rm T} = 32$, $N_{\rm R}  = 32$, $N_{\rm RIS} = 128$, $L = 1$, $T = 1$, $M = 128$, $\Delta f = 120$ KHz. The transmit power and the noise power are set to be $50$~dBm and $-103$~dBm, respectively.}
%	\label{FigADDD}
%\end{figure}

%\begin{figure}[!t]
%	\centering
%	\includegraphics[width=80mm]{ADDomainDulp.pdf}
%	\caption{The geometry relation between the BS, the RIS and the UT}
%	\label{figBSUTRelation}
%\end{figure}
%
%\begin{figure}[!t]
%	\centering
%	\includegraphics[width=80mm]{DDDomainDulp.pdf}
%	\caption{The geometry relation between the BS, the RIS and the UT}
%	\label{figBSUTRelation}
%\end{figure}
%We may use the difference between the properties of the echoes from the RIS and the those of  the echoes from the targets to perform beam training and target detection based on the mixed echoes from the RIS and the targets in \eqref{BeamTrainingStack}.
From \eqref{BeamTraining1DFT} and \eqref{BeamTraining2DFT},  the energy of the echoes from the RIS is accumulated in the angle-delay domain and is dispersed  in the Doppler-delay domain while that from the targets is accumulated in the angle-delay domain and is dispersed in the angle-delay domain. To unveil the inherent difference between the echoes from the RIS and the targets, we analyze the LoS path of the channel between the BS and the RIS  and take the target with the biggest received power as an example at the same time. Define
\begin{align}\label{BeamTraining2DFTCp}
	&\widetilde{\boldsymbol{Y}}_n^{(\rm c)} \triangleq \mu_{1,1}\boldsymbol{F}_{N_{\rm RIS}} \boldsymbol{\alpha}\left(N_{\rm RIS},2\varphi_{\rm BR}^{(1)}\right)\widetilde{\boldsymbol{e}}_{1,1}^{\rm T}, \nonumber \\
	&\widetilde{Q}^{\rm (c)}\triangleq\max_{n,s,m} \left|[\widetilde{\boldsymbol{Y}}_n^{(\rm c)}]_{s,m}\right|, \nonumber \\
	&\widetilde{\boldsymbol{Y}}_n^{(\rm r)} \triangleq \kappa_1 \boldsymbol{\alpha}\left(N_{\rm RIS},2f_{\rm Tar}^{(1)}T_{\rm s}\right)\widetilde{\boldsymbol{d}}_{1}^{\rm T},\widetilde{Q}^{\rm (r)}\triangleq\max_{n,s,m} \left|[\widetilde{\boldsymbol{Y}}_n^{(\rm r)}]_{s,m}\right|,\nonumber \\
	&\overline{\boldsymbol{Y}}_n^{(\rm c)} \triangleq \mu_{1,1}\boldsymbol{\alpha}\left(N_{\rm RIS},2\varphi_{\rm BR}^{(1)}\right)\widetilde{\boldsymbol{e}}_{1,1}^{\rm T},\overline{Q}^{\rm (c)}\triangleq\max_{n,s,m} \left|[\overline{\boldsymbol{Y}}_n^{(\rm c)}]_{s,m}\right|,\nonumber \\
	&\overline{\boldsymbol{Y}}_n^{(\rm r)} \triangleq  \kappa_1\widetilde{\boldsymbol{c}}_1\widetilde{\boldsymbol{d}}_{1}^{\rm T},~~\overline{Q}^{\rm (r)}\triangleq\max_{n,s,m} \left|[\overline{\boldsymbol{Y}}_n^{(\rm r)}]_{s,m}\right|,
\end{align}
where  we specify the first path as the LoS path and the first target as the target with the biggest received power, without loss of generality. Suppose $\widetilde{Q}^{\rm (c)} \ge \overline{Q}^{\rm (r)}$, that  is, the accumulated power of the echoes from the RIS in the angle-delay domain is stronger than the accumulated power of the echoes from the targets in the Doppler-delay domain. Therefore,
\begin{equation}\label{Ratio1}
	\frac{\widetilde{Q}^{\rm (c)}}{\widetilde{Q}^{\rm (r)}} \ge \frac{\widetilde{Q}^{\rm (c)}}{\overline{Q}^{\rm (r)}}\cdot\frac{\overline{Q}^{\rm (r)}}{\widetilde{Q}^{\rm (r)}}\ge 1 \cdot \sqrt{N_{\rm RIS}/2} = \sqrt{N_{\rm RIS}/2}.
\end{equation}
Similarly, if $\overline{Q}^{\rm (r)} > \widetilde{Q}^{\rm (c)}$, we have ${\overline{Q}^{\rm (r)}}/{\overline{Q}^{\rm (c)}}>\sqrt{N_{\rm RIS}/2}$. From \eqref{Ratio1}, if $\widetilde{Q}^{\rm (c)} \ge \overline{Q}^{\rm (r)}$,  the dispersed power of the  echoes from the target is smaller than the accumulated power of the echoes from the  RIS  in the angle-delay domain. On the other hand, if $\widetilde{Q}^{\rm (c)} < \overline{Q}^{\rm (r)}$, the dispersed power of the echoes from the RIS is smaller than the accumulated power of the  echoes from the target in the Doppler-delay domain. Benefiting from the difference between  the RIS and the targets, we can distinguish them even if the targets move at low speeds. To show the accumulation and dispersion of signals  intuitively, we illustrate the signals in the angle-delay domain and the Doppler-delay domain in Fig.~\ref{FigADDD}. As shown in  Fig.~\ref{FigAD}, in the angle-delay domain, the energy of the echoes from the RIS is accumulated while the energy of the echoes from the target is dispersed, and the power of the accumulated signal of the RIS is about 20~dB higher than  the power of the dispersed signal of the target. On the contrary, in the Doppler-delay domain, the energy of the echoes from the RIS is dispersed  while the energy of the echoes from the target is accumulated, and the power of the accumulated signal of the RIS is about 20~dB lower than the power of the dispersed signal of the target. 

Based on the above findings, we then propose the IPEBTTS algorithm to distinguish the RIS from the targets with the mixed echoes from the RIS and the targets. Note that different from the work that is only adapted to the LoS channels~\cite{icc23CKJ}, the IPEBTTS algorithm is adapted to both the LoS channels and NLoS channels.

\subsection{Iterative Parameter Estimation for Beam Training and Target Sensing}
The proposed IPEBTTS algorithm consists of five parts, including initialization, transformation, parameter estimation for beam training, parameter estimation for target sensing and stop condition, which are detailed in the following.  
\subsubsection{Initialization}\label{SecIIIB}
We denote the initial received signals  of the $n$th transmit beam in~\eqref{BeamTrainingStack} as $\boldsymbol{Y}_{n}^{(0)}$, for $n =1,2,\cdots, N_{\rm T}$, and initialize the number of detection operations as $q = 0$.
\subsubsection{Transformation}\label{SecIIIC}
We update $q$ as $q\leftarrow q + 1$. When performing the $q$th detection, the angle-delay domain signals, $\widetilde{\boldsymbol{Y}}_{n}^{(q)}$,  and the Doppler-delay domain signals, $\overline{\boldsymbol{Y}}_{n}^{(q)}$, can be expressed as
\begin{align}\label{ADvsDD}
	& \widetilde{\boldsymbol{Y}}_{n}^{(q)} = \boldsymbol{Y}_{n}^{(q-1)}\boldsymbol{F}_M, \nonumber \\
	& \overline{\boldsymbol{Y}}_{n}^{(q)} = \boldsymbol{W}_{N_{\rm RIS}}^{\rm T}\widetilde{\boldsymbol{Y}}_{n}^{(q)}/\sqrt{N_{\rm RIS}}.
\end{align}
respectively. Denote the maximum values in the angle-delay domain and the Doppler-delay domain as $\widehat{Q}_{\rm c}^{(q)}$ and $\widehat{Q}_{\rm r}^{(q)}$, respectively. We have 
\begin{align}\label{MaxValue}
	&\widehat{Q}_{\rm c}^{(q)} =  \max_{n,s,m} \left|\left[\boldsymbol{\widetilde{Y}}_n^{(q)}\right]_{s,m}\right|,~\widehat{Q}_{\rm r}^{(q)} =  \max_{n,s,m} \left|\left[\boldsymbol{\overline{Y}}_n^{(q)}\right]_{s,m}\right|
\end{align}
for $n = 1,\cdots,N_{\rm T},~ s = 1,\cdots,N_{\rm RIS}$, and $ m = 1,\cdots,M$.
\subsubsection{Parameter Estimation for Beam Training}\label{SecIIID}
 If $\widehat{Q}_{\rm c}^{(q)} \ge \widehat{Q}_{\rm r}^{(q)}$, that is, the maximum accumulated power of the echoes from the RIS is bigger than the maximum accumulated power of the echoes from the targets, we can focus on the angle-delay domain and treat the echoes from the targets as the interference. Then, we can obtain the beam training result of $\boldsymbol{\widetilde{Y}}_n^{(q)}$ via
\begin{align}\label{BeamTrainingResult}
&\left(n^{(q)}_{\rm RIS},s^{(q)}_{\rm RIS},m^{(q)}_{\rm RIS}\right) = \arg\max_{n,s,m} \left|\left[\boldsymbol{\widetilde{Y}}_n^{(q)}\right]_{s,m}\right|
\end{align}
for $n = 1,\cdots,N_{\rm T},~ s = 1,\cdots,N_{\rm RIS},$ and $m = 1,\cdots,M$,  where $n^{(q)}_{\rm RIS}$  and $s^{(q)}_{\rm RIS}$ denote the indices of the beam in $\boldsymbol{\mathcal{B}}$ and $\boldsymbol{\mathcal{R}}$, respectively, and $m^{(q)}_{\rm RIS}$ denotes the index of the column of $\boldsymbol{F}_M$ best fitting for the channel contained in $\boldsymbol{\widetilde{Y}}_n^{(q)}$. Note that the estimation accuracy in \eqref{BeamTrainingResult} is limited by the DFT resolution. To obtain precise channel state information, the accuracy of the beam training results in \eqref{BeamTrainingResult} needs to be improved.

Denote the $N\times N$ DFT matrix as $\boldsymbol{W}_N$. For $\boldsymbol{g} = \boldsymbol{\alpha}(N,\vartheta)$, we define
\begin{align}\label{DFTData}
	&\widetilde{\boldsymbol{g}} \triangleq \boldsymbol{W}_N\boldsymbol{g}, ~~g^* \triangleq \arg\max_{g = 1,\cdots,N} |[\widetilde{\boldsymbol{g}}]_g|, \nonumber \\
	&\widetilde{g}^* \triangleq \arg\max_{g = 1,\cdots,N,g\neq g^*} |[\widetilde{\boldsymbol{g}}]_g|.
\end{align}
Then, the estimation of $\vartheta$ can be expressed as~\cite{TWC17ZDL}
\begin{align}\label{AngleEquations}
	\widehat{\vartheta }&=\left( -1+\frac{\widetilde{g}^*+g^*}{N} \right) \nonumber\\
	&-\frac{1}{\pi}\mathrm{asin} \left( \frac{\Gamma \sin\mathrm{(}\delta )-\Gamma \sqrt{1-\Gamma ^2}\sin\mathrm{(}\delta )\cos\mathrm{(}\delta )}{\sin\mathrm{(}\delta )^2+\Gamma ^2\cos\mathrm{(}\delta )^2} \right),
\end{align}
where
\begin{equation}\label{GammaEquation}
	\Gamma = {\rm sgn}(\widetilde{g}^* - g^* )\frac{|[\widetilde{\boldsymbol{g}}]_{g^*}|^2 -|[\widetilde{\boldsymbol{g}}]_{\widetilde{g}^*}|^2}{|[\widetilde{\boldsymbol{g}}]_{g^*}|^2 +|[\widetilde{\boldsymbol{g}}]_{\widetilde{g}^*}|^2},
\end{equation}
and $\delta = \pi/N$. Based on \eqref{DFTData} to \eqref{GammaEquation}, the angle of a steering vector can be estimated with the aid of the DFT/IDFT. 
 
Note that both the beam sweeping at the RIS and the subcarrier accumulation in \eqref{BeamTraining2DFT} are the DFT. After substituting 	$N \leftarrow N_{\rm RIS}$ and $\widetilde{\boldsymbol{g}} \leftarrow \big[\boldsymbol{\widetilde{Y}}_{n^{(q)}_{\rm RIS}}^{(q)}\big]_{:,m^{(q)}_{\rm RIS}}$ into \eqref{DFTData}, \eqref{AngleEquations} and \eqref{GammaEquation}, we denote the result of \eqref{AngleEquations} as $\widetilde{\vartheta}_q$. According to \eqref{BeamTrainingStack}, $\widetilde{\vartheta}_q$ is the sum of the two-way AoAs. Therefore, we have 
\begin{equation}\label{TwoWayAoA}
\widetilde{\vartheta}_q = 2\widehat{\varphi}_{\rm BR}^{(q)} + 2\widetilde{n},
\end{equation}
for $\widetilde{n}\in\mathbb{N}$, where $\widehat{\varphi}_{\rm BR}^{(q)}$ is an estimate of the AoA  best fitting for the channel contained in  $\boldsymbol{\widetilde{Y}}_n^{(q)}$. Since $\widehat{\varphi}_{\rm BR}^{(q)}\in [-1,1]$ and $\widetilde{\vartheta}_q \in [-1,1]$, we have 
\begin{equation}\label{OneWayAoA}
\widehat{\varphi}_{\rm BR}^{(q)} = \begin{cases}
\widetilde{\vartheta }_q/2 + 1~\mathrm{or}~\widetilde{\vartheta }_q/2,~\widetilde{\vartheta}_q\le 0,\\
	\widetilde{\vartheta }_q/2 - 1~\mathrm{or}~\widetilde{\vartheta }_q/2,~\widetilde{\vartheta}_q>0.\\
\end{cases}
\end{equation}
Denote  set $\boldsymbol{\varUpsilon}_{\rm BR}^{(q)}$ to be all the possible values of $\widehat{\varphi}_{\rm BR}^{(q)}$. Then, we have 
\begin{equation}\label{OneWayAoAset}
	\boldsymbol{\varUpsilon}_{\rm BR}^{(q)} = \begin{cases}
		\{\widetilde{\vartheta }_q/2 + 1,~\widetilde{\vartheta }_q/2\},~\widetilde{\vartheta}_q\le 0,\\
		\{\widetilde{\vartheta }_q/2 - 1,~\widetilde{\vartheta }_q/2\},~\widetilde{\vartheta}_q>0.\\
	\end{cases}
\end{equation}
Due to the periodicity of $\widehat{\varphi}_{\rm BR}^{(q)}$, it is difficult to select the correct value from the two candidates, which will be addressed in detail in Section~\ref{SBA}. 
%In \eqref{OneWayAoA}, for each $\widetilde{\vartheta}_q$, two estimations of $\widehat{\varphi}_{\rm BR}^{(q)}$ can be obtained. 

Denote the output of \eqref{AngleEquations} as $\psi_q$ after substituting  $N \leftarrow M$ and $\widetilde{\boldsymbol{g}} \leftarrow \big[\boldsymbol{\widetilde{Y}}_{n^{(q)}_{\rm RIS}}^{(q){\rm T}}\big]_{:,s_{\rm RIS}^{(q)}}$ into \eqref{DFTData}, \eqref{AngleEquations} and \eqref{GammaEquation}. Then, the estimations of the delay and the distance between the transceiver and the RIS can be expressed as 
\begin{equation}\label{tauBR}
\widehat{\tau}_{\rm BR}^{(q)} = \frac{1 + \psi_q - 2/M}{4\Delta f},~~~\widehat{r}_{\rm BR}^{(q)} = \widehat{\tau}_{\rm BR}^{(q)} c. 
\end{equation}

Different from the estimations of the range and the channel AoA, the estimation of the channel AoD between the BS and the RIS cannot be performed with the equations from \eqref{DFTData} to \eqref{GammaEquation} because the received signals are  concurrently related to both the transmit beams and the receive beams. Alternatively, we estimate AoD in  \eqref{BeamTrainingResult} via a least-squares approach, which can be expressed as 
\begin{align}\label{leastsquare}
	&\min_{\theta,\alpha} \left\|\boldsymbol{y}^{(q)} - \alpha \widetilde{\boldsymbol{y}}^{(q)}   \right\|_2 \nonumber \\
	&{\rm ~s.t.}~\theta \in \left[(2n_{\rm RIS}^{(q)}-3)/N_{\rm T},(2n_{\rm RIS}^{(q)}-1)/N_{\rm T}\right]
\end{align}
where 
\begin{align}\label{leastvar}
	&~~~~~~~~~~~~~\boldsymbol{y}^{(q)} = \left[ \begin{array}{c}
		\big[\boldsymbol{\widetilde{Y}}_{n^{(q)}_{\rm RIS}-1}^{(q)}\big]_{s_{\rm RIS}^{(q)},m_{\rm RIS}^{(q)}} \\
		\big[\boldsymbol{\widetilde{Y}}_{n^{(q)}_{\rm RIS}}^{(q)}\big]_{s_{\rm RIS}^{(q)},m_{\rm RIS}^{(q)}} \\
		\big[\boldsymbol{\widetilde{Y}}_{n^{(q)}_{\rm RIS}+1}^{(q)}\big]_{s_{\rm RIS}^{(q)},m_{\rm RIS}^{(q)}} \\
	\end{array} \right] \nonumber \\
	&\widetilde{\boldsymbol{y}}^{(q)} = \left[ \begin{array}{c}
		\boldsymbol{v}_{n^{(q)}_{\rm RIS}-1}^{\rm H} \boldsymbol{\alpha}\left(N_{\rm R},\theta\right) \boldsymbol{\alpha}\left(N_{\rm T},\theta\right)^{\rm H}\boldsymbol{b}_{n^{(q)}_{\rm RIS}-1} \\
		\boldsymbol{v}_{n^{(q)}_{\rm RIS}}^{\rm H} \boldsymbol{\alpha}\left(N_{\rm R},\theta\right) \boldsymbol{\alpha}\left(N_{\rm T},\theta\right)^{\rm H}\boldsymbol{b}_{n^{(q)}_{\rm RIS}} \\
		\boldsymbol{v}_{n^{(q)}_{\rm RIS}+1}^{\rm H} \boldsymbol{\alpha}\left(N_{\rm R},\theta\right) \boldsymbol{\alpha}\left(N_{\rm T},\theta\right)^{\rm H}\boldsymbol{b}_{n^{(q)}_{\rm RIS}+1} \\
	\end{array} \right]. 
\end{align}
For a fixed $\theta$, the optimal $\alpha$ to \eqref{leastsquare} can be expressed as $\widehat{\alpha} = \widetilde{\boldsymbol{y}}^{(q){\rm H}}\boldsymbol{y}^{(q)}/ \widetilde{\boldsymbol{y}}^{(q){\rm H}}\widetilde{\boldsymbol{y}}^{(q)}$. Then, \eqref{leastsquare} can be rewritten to 
\begin{align}\label{leastsquare2}
	&\max_{\theta} \big\|\boldsymbol{y}^{(q){\rm H}}\widetilde{\boldsymbol{y}}^{(q)}\big\|_2\big/\big\|\widetilde{\boldsymbol{y}}^{(q)}\big\|_2 \nonumber \\
	&{\rm ~s.t.}~\theta \in \left[(2n_{\rm RIS}^{(q)}-3)/N_{\rm T},(2n_{\rm RIS}^{(q)}-1)/N_{\rm T}\right].
\end{align}
Since \eqref{leastsquare2} is  a single-variable optimization problem with respect to bounded variable $\theta$, it can be solved with numerical techniques. We omit the detailed procedure and  denote  the solution of \eqref{leastsquare2} as $\widehat{\theta}_{\rm BR}^{(q)}$.

After estimating one of the channel paths, we need to remove the estimated contributions from the received signals to facilitate the estimation of other channel paths or targets.  The estimated contributions can be removed from the received signals via
\begin{align}\label{RISRemoval}
&\min_{\beta} \big\|\overline{\boldsymbol{y}}^{(q)} - \beta\boldsymbol{t}\big\|_2, 
\end{align}
where
\begin{align}\label{RISRemovalVar}
	&\overline{\boldsymbol{y}}^{(q)} = \left[{\rm vec}\left(\boldsymbol{{Y}}_1^{(q-1)}\right)^{\rm T},\cdots,{\rm vec}\left(\boldsymbol{{Y}}_{N_{\rm T}}^{(q-1)}\right)^{\rm T}\right]^{\rm T} \nonumber\\
	&\boldsymbol{t} = \widetilde{\boldsymbol{t}} \otimes\boldsymbol{\alpha}\left(M,-2\widehat{\tau}_{\rm BR}^{(q)}\Delta f\right)  \otimes \left(\boldsymbol{F}_{N_{\rm RIS}}\boldsymbol{\alpha}\left(N_{\rm RIS},\widetilde{\vartheta}_q\right)\right)\nonumber \\
	&\widetilde{\boldsymbol{t}} = \left[ \begin{array}{c}
		\boldsymbol{v}_{1}^{\rm H} \boldsymbol{\alpha}\left(N_{\rm R},\widehat{\theta}_{\rm BR}^{(q)}\right) \boldsymbol{\alpha}\left(N_{\rm T},\widehat{\theta}_{\rm BR}^{(q)}\right)^{\rm H}\boldsymbol{b}_{1} \\
		\vdots  \\
		\boldsymbol{v}_{N_{\rm RIS}}^{\rm H} \boldsymbol{\alpha}\left(N_{\rm R},\widehat{\theta}_{\rm BR}^{(q)}\right) \boldsymbol{\alpha}\left(N_{\rm T},\widehat{\theta}_{\rm BR}^{(q)}\right)^{\rm H}\boldsymbol{b}_{N_{\rm RIS}} \\
	\end{array} \right].
\end{align}
The optimal $\beta$ to \eqref{RISRemoval} can be expressed as $\widehat{\beta} = \boldsymbol{t}^{{\rm H}}\overline{\boldsymbol{y}}^{(q)}/ \boldsymbol{t}^{{\rm H}}\boldsymbol{t}$. Comparing \eqref{BeamTrainingStack} and \eqref{RISRemoval}, we have $\sqrt{M}\widehat{g}_{\rm BR}^{(q)}\widehat{g}_{\rm BR}^{(q)} = \widehat{\beta}$. As a result, the  channel factor of the detected channel path can be estimated via 
\begin{align}\label{ChannelFactor}
\widehat{g}_{\rm BR}^{(q)} = \sqrt{\widehat{\beta}/\sqrt{M}}.
\end{align}
  Then, we obtain $\boldsymbol{Y}_{n}^{(q)}$ via 
\begin{align}\label{BTYnq}
	\boldsymbol{Y}_{n}^{(q)} &= \boldsymbol{Y}_{n}^{(q-1)}\nonumber \\ 
	&- \widehat{\beta}[\widetilde{\boldsymbol{t}}]_{n}\boldsymbol{F}_{N_{\rm RIS}}\boldsymbol{\alpha}(N_{\rm RIS},\widetilde{\vartheta}_q)\boldsymbol{\alpha}(M,-2\widehat{\tau}_{\rm BR}^{(q)}\Delta f)^{\rm T}.
\end{align}
\subsubsection{Parameter Estimation for  Target Sensing}\label{SecIIIE}
%that is the maximum accumulated power of the echoes from the targets is bigger than the maximum accumulated power of the echoes from the RIS, we can focus on the Doppler-delay domain and call the echoes from the RIS as interference according to \eqref{Ratio1}.
If $\widehat{Q}_{\rm c}^{(q)} < \widehat{Q}_{\rm r}^{(q)}$,  the results of the target sensing can be obtained via
\begin{align}\label{BeamTrainingResultTar}
	\left(n^{(q)}_{\rm Tar},s^{(q)}_{\rm Tar},m^{(q)}_{\rm Tar}\right) = \arg\max_{n,s,m} \left|\big[\overline{\boldsymbol{Y}}_{n}^{(q)}\big]_{s,m}\right|.
\end{align}
%Denote the output of \eqref{AngleEquations} as $\widetilde{\psi}_q$ after substituting $N \leftarrow N_{\rm RIS}$ and $\widetilde{\boldsymbol{g}} \leftarrow \big[\boldsymbol{\overline{Y}}_{n^{(q)}_{\rm Tar}}^{(q)}\big]_{:,m_{\rm Tar}^{(q)}}$ into \eqref{DFTData}, \eqref{AngleEquations} and \eqref{GammaEquation}. Then, the Doppler frequency and the velocity of the target can be expressed as
%\begin{equation}\label{dpTar}
%\widehat{f}_{\rm Tar}^{(q)}  = \frac{1 + \widetilde{\psi}_q - 2/N_{\rm RIS}}{2T_{\rm s}},~~~\widehat{v}_{\rm Tar}^{(q)} = \widehat{f}_{\rm Tar}^{(q)} \lambda/2. 
%\end{equation} 
%After substituting $N \leftarrow M$ and $\widetilde{\boldsymbol{g}} \leftarrow \big[\boldsymbol{\overline{Y}}_{n^{(q)}_{\rm Tar}}^{(q){\rm T}}\big]_{:,s_{\rm Tar}^{(q)}}$ into \eqref{DFTData}, \eqref{AngleEquations} and \eqref{GammaEquation}, we denote the output of \eqref{AngleEquations} as $\widetilde{\psi}_q$. Then, the delay and the distance of the target can be expressed as
%\begin{equation}\label{tauTar}
%	\widehat{\tau}_{\rm Tar}^{(q)} = \frac{1 + \widetilde{\psi}_q - 2/M}{4\Delta f},~~~\widehat{r}_{\rm Tar}^{(q)} = \widehat{\tau}_{\rm Tar}^{(q)} c. 
%\end{equation}
%Similar to \eqref{leastsquare}, the estimated AoD of the target can also be  obtained by a least square approach, which is denoted as $\widehat{\theta}_{\rm Tar}^{(q)}$. 
Note that signals in the Doppler-delay domain are similar to those in the angle-delay domain. We can obtain $\widehat{v}_{\rm Tar}^{(q)}$, $\widehat{r}_{\rm Tar}^{(q)}$, $\widehat{\theta}_{\rm Tar}^{(q)}$ and $\boldsymbol{Y}_{n}^{(q)}$ according to \eqref{tauBR}, \eqref{tauBR}, \eqref{leastsquare2} and \eqref{BTYnq}, respectively. In addition, the parameters of the targets are estimated based on three adjacent transmit codewords. Therefore, we can also perform \textit{Parameter Estimation for Target Sensing} every $3N_{\rm RIS}$ time slots to reduce the delay of target sensing.
%Then, we remove the contributions of the estimated target via
%\begin{equation}\label{projection2}
%\min_{\varUpsilon} \big\|\boldsymbol{y}^{(q)} - \varUpsilon \boldsymbol{s}\big\|_2 
%\end{equation}
%where
%\begin{align}
%	&\boldsymbol{s} = \widetilde{\boldsymbol{s}} \otimes \boldsymbol{\alpha}\left(M,-2\widehat{\tau}_{\rm Tar}^{(q)}\Delta f\right) \otimes \boldsymbol{\alpha }\left(N_{\mathrm{RIS}},2\pi \widehat{f}_{\rm Tar}^{(q)}T_s\right), \nonumber\\
%	&\widetilde{\boldsymbol{s}} = \left[ \begin{array}{c}
%		\boldsymbol{v}_{1}^{\rm H} \boldsymbol{\alpha}\left(N_{\rm R},\widehat{\theta}_{\rm Tar}^{(q)}\right) \boldsymbol{\alpha}\left(N_{\rm T},\widehat{\theta}_{\rm Tar}^{(q)}\right)^{\rm H}\boldsymbol{b}_{1} \\
%		\vdots  \\
%		\boldsymbol{v}_{N_{\rm RIS}}^{\rm H} \boldsymbol{\alpha}\left(N_{\rm R},\widehat{\theta}_{\rm Tar}^{(q)}\right) \boldsymbol{\alpha}\left(N_{\rm T},\widehat{\theta}_{\rm Tar}^{(q)}\right)^{\rm H}\boldsymbol{b}_{N_{\rm RIS}} \\
%	\end{array} \right].
%\end{align}
%The optimal solution to \eqref{projection2} is $\widehat{\varUpsilon} = \boldsymbol{s}^{{\rm H}}\boldsymbol{y}^{(q)}/ \boldsymbol{s}^{{\rm H}}\boldsymbol{s}$.  Then, we obtain $\boldsymbol{Y}_{n}^{(q)}$ via 
%\begin{align}\label{SenYnq}
%	&\boldsymbol{Y}_{n}^{(q)} = \boldsymbol{Y}_{n}^{(q-1)}- \widehat{\varUpsilon}[\widetilde{\boldsymbol{s}}]_{n}\boldsymbol{\alpha }(N_{\mathrm{RIS}},2\pi \widehat{f}_{\rm Tar}^{(q)}T_s)\boldsymbol{\alpha}(M,-2\widehat{\tau}_{\rm Tar}^{(q)}\Delta f)^{\rm T}.
%\end{align}
\subsubsection{Stop Condition}\label{SecIIIF}
We repeat Section \ref{SecIIIC}, Section \ref{SecIIID} and Section \ref{SecIIIE} until the detected number of channel paths equals the predefined maximum number of channel paths $\widehat{L}_{\rm BR}$ and the detected number of targets equals the predefined maximum number of targets $\widehat{T}$.

%the condition  that the accumulated power of the echoes from the target is smaller than a threshold $\varrho_{\rm r}$ and the condition  that the accumulated power of the  echoes from the RIS is smaller than a threshold $\varrho_{\rm c}$ are both satisfied. Note that $\varrho_{\rm r}$ and $\varrho_{\rm c}$ can be determined referring to the criterion of the constant false alarm rate (CFAR) detector ~\cite{FRSP}. 
%\subsubsection{Summary}\label{SecIIIG}
%%Suppose $\widehat{L}_{\rm BR}$ channel paths and $\widehat{T}$ targets are finally detected via performing the low-complexity IPEBTTS algorithm. 
%The estimated AoDs, AoAs and the ranges of the paths between the BS and the RIS are denote as $\overline{\theta}_{\rm BR}^{(u)}$, $\overline{\varphi}_{\rm BR}^{(u)}$ and $\overline{r}_{\rm BR}^{(u)}$, respectively, for $u = 1,2,\cdots \widehat{L}_{\rm BR}$. The estimated angles, ranges and velocities of the targets are denoted as  $\overline{\theta}_{\rm Tar}^{(l)}$,  $\overline{r}_{\rm Tar}^{(l)}$ and $\overline{v}_{\rm Tar}^{(l)}$, respectively, for $l=1,\cdots,\widehat{T}$. 
\subsection{Beam Training for the BS-UT Link}
Note that the UT can simultaneously receive the downlink pilots during the beam training between the BS and the RIS. Reusing the transmitted pilots in \eqref{BeamTrainingModel}, the received signals at the UT can be expressed as
\begin{equation}\label{SystemModelofUT2}
	{z}_{m}^{(p)} = \boldsymbol{u}_{t}^{\rm H} (\boldsymbol{H}_{{\rm c},m} + \widetilde{\boldsymbol{H}}_{{\rm c},m}\boldsymbol{\Phi}_p\boldsymbol{G}_{{\rm T},m}) \boldsymbol{b}_{n} + \boldsymbol{u}_{t}^{\rm H}\boldsymbol{\eta}_{{\rm c},m}^{(p)},
\end{equation}
where $n \in \{1,2,\cdots N_{\rm T}\} $, $t \in \{1,2,\cdots,N_{\rm UT}\}$ and $p = (n-1)N_{\rm RIS} + t$.
% In most cases, the channel gain of the BS-RIS-UT link is much smaller than the channel gain of the BS-UT link. This is because the former relies on the reflection of the RIS where a large part of energy is lost after reflection, while the latter relies on the direct propagation which avoids the energy lost caused by reflection. As a result, we can take the BS-RIS-UT link as a NLoS path between the BS and the UT and perform beam training with the interference from the RIS ignored. However, with the increase of number of RIS elements, the assumption that the channel gain of the BS-RIS-UT link is much smaller than the channel gain of the BS-UT link might not hold  because of the high beam gain provided by the RIS. To avoid the confusion between the BS-UT link and the BS-RIS-UT link, we propose a BS-UT link beam training scheme by reusing the training pilots for the BS-RIS-BS link. 
 Suppose $N_{\rm RIS}\geq N_{\rm UT}+2$, we have the following two findings: 1) For each transmit beam, only $N_{\rm UT}$ out of $N_{\rm RIS}$ training symbols are used to perform the beam training between the BS and the UT. 2) The received signals at the UT contain both the signals from the BS and the  signals reflected from the RIS. The first finding inspires us to  use the remaining $N_{\rm RIS} - N_{\rm UT}$ training symbols and the second one inspires us to find out whether the training result corresponds to the direct path from the BS or the reflected path from the RIS. In the following, we will try to utilize the remaining training pilots to distinguish the direct path from the reflected~path.

Suppose $\widehat{n}_{\rm UT}$, $\widehat{t}_{\rm UT}$ and $\widehat{p}_{\rm UT}$ denote the index of the best codeword in $\boldsymbol{\mathcal{B}}$, the index of the best codeword in $\boldsymbol{\mathcal{U}}$ and the corresponding index of training symbol, respectively. If $\widehat{n}_{\rm UT}$, $\widehat{t}_{\rm UT}$ and $\widehat{p}_{\rm UT}$ are the training results for the reflected path from the RIS, then we have
\begin{equation}\label{SystemModelofUT21}
	{z}_{m}^{(\widehat{p}_{\rm UT})} \approx \boldsymbol{u}_{\widehat{t}_{\rm UT}}^{\rm H}\widetilde{\boldsymbol{H}}_{{\rm c},m}\boldsymbol{\Phi}_{\widehat{p}_{\rm UT}}\boldsymbol{G}_{{\rm T},m} \boldsymbol{b}_{\widehat{n}_{\rm UT}} + \boldsymbol{u}_{\widehat{t}_{\rm UT}}^{\rm H} \boldsymbol{\eta}_{{\rm c},m}^{(\widehat{p}_{\rm UT})},
\end{equation}
where we omit the channel from the BS due to the low probability of the simultaneous beam alignment for both the BS-RIS-UT link and BS-UT link. For the $(\widehat{p}_{\rm UT} + 2)$th transmit symbol, the transceiver still transmits the pilot with $\boldsymbol{b}_{\widehat{n}_{\rm UT}}$ according to \eqref{BeamTrainingModel} and the UT still receives the pilot with $\boldsymbol{u}_{\widehat{t}_{\rm UT}}$  while the RIS reflects the pilot with  $\boldsymbol{\Phi}_{\widehat{p}_{\rm UT} + 2}$. Then, the received signal at the UT can be expressed as
\begin{equation}\label{SystemModelofUT22}
	{z}_{m}^{(\widehat{p}_{\rm UT}+2)} \approx \boldsymbol{u}_{\widehat{t}_{\rm UT}}^{\rm H}\widetilde{\boldsymbol{H}}_{{\rm c},m}\boldsymbol{\Phi}_{\widehat{p}_{\rm UT} + 2}\boldsymbol{G}_{{\rm T},m} \boldsymbol{b}_{\widehat{n}_{\rm UT}} + \boldsymbol{u}_{\widehat{t}_{\rm UT}}^{\rm H}\boldsymbol{\eta}_{{\rm c},m}^{(\widehat{p}_{\rm UT}+2)}.
\end{equation}	
Based on \eqref{SystemModelofUT21} and \eqref{SystemModelofUT22}, we have
\begin{equation}\label{ratio1}
	\rho = \big|{z}_{m}^{(\widehat{p}_{\rm UT}+2)}\big|/\big|{z}_{m}^{(\widehat{p}_{\rm UT})}\big| \approx 0,
\end{equation}
due to the power difference between the beam alignment and beam misalignment.

 On the other hand, if $\widehat{n}_{\rm UT}$, $\widehat{t}_{\rm UT}$ and $\widehat{p}_{\rm UT}$ are the training results for the direct path, we have
\begin{align}\label{SystemModelofUT4}
	&{z}_{m}^{(\widehat{p}_{\rm UT})} \approx \boldsymbol{u}_{\widehat{t}_{\rm UT}}^{\rm H}\boldsymbol{H}_{{\rm c},m}\boldsymbol{b}_{\widehat{n}_{\rm UT}} + \boldsymbol{u}_{\widehat{t}_{\rm UT}}^{\rm H}\boldsymbol{\eta}_{{\rm c},m}^{(\widehat{p}_{\rm UT})}, \nonumber\\
	&{z}_{m}^{(\widehat{p}_{\rm UT}+2)} \approx \boldsymbol{u}_{\widehat{t}_{\rm UT}}^{\rm H}\boldsymbol{H}_{{\rm c},m}\boldsymbol{b}_{\widehat{n}_{\rm UT}} + \boldsymbol{u}_{\widehat{t}_{\rm UT}}^{\rm H}\boldsymbol{\eta}_{{\rm c},m}^{(\widehat{p}_{\rm UT}+2)}. \\
	&\rho = \big|{z}_{m}^{(\widehat{p}_{\rm UT}+2)}\big|/\big|{z}_{m}^{(\widehat{p}_{\rm UT})}\big| \approx 1. \label{ratio2}
\end{align}	
Note that the value of $\rho$ in \eqref{ratio2} differs from that in \eqref{ratio1} because the beam training in \eqref{BeamTrainingModel} is dedicated to the BS-RIS-BS link and the BS-UT link instead of the BS-RIS-UT link. Exploiting the difference between \eqref{ratio1} and \eqref{ratio2}, we only need two more time slots to identify whether the training results correspond to the direct path from the BS or the reflected path from the RIS. In the following, we elaborate the BS-UT link beam training by reusing the training pilots for the BS-RIS-BS link. 
\begin{algorithm}[!t]
	\caption{Simultaneous Beam Training and Target Sensing (SBTTS) Scheme}
	\label{alg1}
	\begin{algorithmic}[1]
		\STATE \textbf{Input:} $N_{\rm T}$, $N_{\rm R}$, $N_{\rm RIS}$, $N_{\rm UT}$, $M$, $\Delta f$, $T_{\rm s}$, $\boldsymbol{\mathcal{B}}$, $\boldsymbol{\mathcal{U}}$, $\boldsymbol{\mathcal{R}}$.
		\STATE \textbf{Initialization:} $q\leftarrow 0$, $u\leftarrow 0$, $l\leftarrow 0$, $k\leftarrow 0$. 
		\STATE ~~~~~~~~~~~~~~~~~~Obtain $\boldsymbol{Y}_{n}^{(0)}$ for $n =1,\cdots N_{\rm T}$ via \eqref{BeamTrainingStack}.
		\STATE ~~~~~~~~~~~~~~~~~~Obtain $\boldsymbol{Z}_{n}^{(0)}$ for $n =1,\cdots N_{\rm T}$ via \eqref{SystemModelofUT2}.
		\WHILE{$u\le \widehat{L}_{\rm BR}$ and $l\le \widehat{T}$}
		\STATE $q\leftarrow q + 1$. 
		\STATE Obtain $\widetilde{\boldsymbol{Y}}_{n}^{(q)}$ and $\overline{\boldsymbol{Y}}_{n}^{(q)}$ via \eqref{ADvsDD}.
		\STATE Obtain $\widehat{Q}_{\rm c}^{(q)}$ and $\widehat{Q}_{\rm r}^{(q)}$ via \eqref{MaxValue}.   
		\IF{$\widehat{Q}_{\rm c}^{(q)} \ge \widehat{Q}_{\rm r}^{(q)}$}
		\STATE Obtain $\boldsymbol{\varUpsilon}_{\rm BR}^{(q)}$, $\widehat{r}_{\rm BR}^{(q)}$, $\widehat{\theta}_{\rm BR}^{(q)}$, $\widehat{g}_{\rm BR}^{(q)}$  and $\boldsymbol{Y}_{n}^{(q)}$ via \eqref{OneWayAoAset}, \eqref{tauBR}, \eqref{leastsquare2}, \eqref{ChannelFactor} and \eqref{BTYnq}, respectively.
		\STATE $u\leftarrow u +1 $.
		\STATE $\overline{\boldsymbol{\varUpsilon}}_{\rm BR}^{(u)}\!\leftarrow\!\boldsymbol{\varUpsilon}_{\rm BR}^{(q)}$,  $\overline{r}_{\rm BR}^{(u)} \!\leftarrow\!\widehat{r}_{\rm BR}^{(q)}$, $\overline{\theta}_{\rm BR}^{(u)}\!\leftarrow\!\widehat{\theta}_{\rm BR}^{(q)}$, $\overline{g}_{\rm BR}^{(u)}\!\leftarrow\!\widehat{g}_{\rm BR}^{(q)}$.
		\ELSE
		\STATE Obtain $\widehat{v}_{\rm Tar}^{(q)}$, $\widehat{r}_{\rm Tar}^{(q)}$, $\widehat{\theta}_{\rm Tar}^{(q)}$ and $\boldsymbol{Y}_{n}^{(q)}$ according to \eqref{tauBR}, \eqref{tauBR}, \eqref{leastsquare2} and \eqref{RISRemoval}, respectively.
		\STATE $l\leftarrow l + 1$.
		\STATE $\overline{v}_{\rm Tar}^{(l)} \leftarrow \widehat{v}_{\rm Tar}^{(q)}$,~$\overline{r}_{\rm Tar}^{(l)} \leftarrow \widehat{r}_{\rm Tar}^{(q)}$, and~$\overline{\theta}_{\rm Tar}^{(l)} \leftarrow \widehat{\theta}_{\rm Tar}^{(q)}$.
		\ENDIF
		\ENDWHILE
%		\STATE $\widehat{L}_{\rm BR} \leftarrow u$,~$\widehat{T} \leftarrow l$. /*Number of paths and targets*/
		\WHILE{ $k\le \widehat{L}_{\rm BU}$}
		\STATE $k\leftarrow k+1 $. 
		\STATE Obtain $\overline{\theta}_{\rm BU}^{(k)}$, $\overline{\varphi}_{\rm BU}^{(k)}$, $\overline{r}_{\rm BU}^{(k)}$ according to \eqref{tauBR}.
		\STATE Obtain $\boldsymbol{Z}_{n}^{(k)}$ according to \eqref{RISRemoval}. 
		\ENDWHILE
%		\STATE $\widehat{L}_{\rm BU} \leftarrow k$. /*Number of paths*/
		\STATE \textbf{Output:} $\overline{\theta}_{\rm BR}^{(u)}$, $\overline{\boldsymbol{\varUpsilon}}_{\rm BR}^{(u)}$, $\overline{r}_{\rm BR}^{(u)}$ and $\widehat{g}_{\rm BR}^{(u)}$ for $u = 1,2,\cdots \widehat{L}_{\rm BR}$.\\
		~~~~~ ~~~~~$\overline{\theta}_{\rm BU}^{(k)}$, $\overline{\varphi}_{\rm BU}^{(k)}$ and $\overline{r}_{\rm BU}^{(k)}$ for $k=1,2,\cdots,\widehat{L}_{\rm BU}$.\\
		~~~~~ ~~~~~$\overline{\theta}_{\rm Tar}^{(l)}$,  $\overline{r}_{\rm Tar}^{(l)}$ and $\overline{v}_{\rm Tar}^{(l)}$, for $l=1,\cdots,\widehat{T}$.
	\end{algorithmic}
\end{algorithm}

\subsubsection{Initialization}\label{inital}
We initialize the number of detection operations as $k=0$. For the $n$th transmit beam, we stack the received signals at the $M$ subcarriers for the $N_{\rm UT}$ receive codewords together as $\boldsymbol{Z}_n^{(0)} \in \mathbb{C}^{N_{\rm UT}\times M}$. We define $\boldsymbol{\Gamma}_{\rm AoD}$, $\boldsymbol{\Gamma}_{\rm AoA}$ and $\boldsymbol{\Gamma}_{\rm Delay}$ to keep the indices of codewords in $\boldsymbol{\mathcal{B}}$, the indices of codewords in $\boldsymbol{\mathcal{U}}$ and the indices of the delay taps for the paths reflected from the RIS, respectively, where $\boldsymbol{\Gamma}_{\rm AoD}$, $\boldsymbol{\Gamma}_{\rm AoA}$ and $\boldsymbol{\Gamma}_{\rm Delay}$ are initially set to be empty.
\subsubsection{Path Detection}\label{UTPathDete}
When performing  the $k$th detection for $k\ge1$, we postmultiply $\boldsymbol{Z}_{n}^{(k-1)}$ with $\boldsymbol{F}_{\rm M}$ for subcarrier accumulation, and obtain $\widetilde{\boldsymbol{Z}}_n^{(k)} = \boldsymbol{Z}_n^{(k-1)}\boldsymbol{F}_{\rm M}$. Then, the training results are obtained via
\begin{align}\label{BeamTrainingResultUT}
	\left(\widehat{n}_{\rm UT}^{(k)}, \widehat{t}_{\rm UT}^{(k)},\widehat{m}_{\rm UT}^{(k)}\right) = \arg\max_{n,t,m} \left|\left[\widetilde{\boldsymbol{Z}}_n^{(k)}\right]_{t,m}\right|.
\end{align}
If $\rho$ calculated from \eqref{ratio1} or \eqref{ratio2} is bigger than the predefined threshold, such as $0.5$, then $ \overline{n}_{\rm UT}^{(k)} = \widehat{n}_{\rm UT}^{(k)}$, $\overline{t}_{\rm UT}^{(k)} = \widehat{t}_{\rm UT}^{(k)}$ and $\overline{m}_{\rm UT}^{(k)} = \widehat{m}_{\rm UT}^{(k)}$ are taken as the training results for the BS-UT link. Otherwise, we update $\boldsymbol{\Gamma}_{\rm AoD}$, $\boldsymbol{\Gamma}_{\rm AoA}$ and $\boldsymbol{\Gamma}_{\rm Delay}$ via 
\begin{align}
	&\boldsymbol{\Gamma}_{\rm AoD} = \boldsymbol{\Gamma}_{\rm AoD} \cup \{\widehat{n}_{\rm UT}^{(k)}-1,\widehat{n}_{\rm UT}^{(k)},\widehat{n}_{\rm UT}^{(k)}+1\}, \nonumber \\
	&\boldsymbol{\Gamma}_{\rm AoA} = \boldsymbol{\Gamma}_{\rm AoA} \cup \{\widehat{t}_{\rm UT}^{(k)}-1,\widehat{t}_{\rm UT}^{(k)},\widehat{t}_{\rm UT}^{(k)}+1\}, \nonumber \\
	&\boldsymbol{\Gamma}_{\rm Delay} = \boldsymbol{\Gamma}_{\rm Delay} \cup \{\widehat{m}_{\rm UT}^{(k)}-1,\widehat{m}_{\rm UT}^{(k)},\widehat{m}_{\rm UT}^{(k)}+1\}. 
\end{align}
Then, we refind $\overline{n}_{\rm UT}^{(k)}$, $\overline{t}_{\rm UT}^{(k)}$ and $\overline{m}_{\rm UT}^{(k)}$ via
\begin{align}\label{BeamTrainingResultUT2}
	&\left(\overline{n}_{\rm UT}^{(k)},\overline{t}_{\rm UT}^{(k)},\overline{m}_{\rm UT}^{(k)}\right) = \arg\max_{n,t,m}  \left|\left[\widetilde{\boldsymbol{Z}}_n^{(k)}\right]_{t,m}\right| \nonumber\\
	&~{\rm s.t.}~n=1,\cdots N_{\rm T}, n\notin\boldsymbol{\Gamma}_{\rm AoD},  \nonumber\\
	&~~~~~~t=1,\cdots N_{\rm UT} ,t\notin\boldsymbol{\Gamma}_{\rm AoA},  \nonumber\\
	&~~~~~~m=1,\cdots M, m\notin\boldsymbol{\Gamma}_{\rm Delay}.
\end{align}
After obtaining the beam training results for the BS-UT link, we can obtain $\overline{\theta}_{\rm BU}^{(k)}$, $\overline{\varphi}_{\rm BU}^{(k)}$, $\overline{r}_{\rm BU}^{(k)}$ according to \eqref{tauBR}. Then, we update $\boldsymbol{Z}_{n}^{(k)}$ according to \eqref{RISRemoval}.
\subsubsection{Stop Condition}\label{UTSC}
We repeat Section \ref{UTPathDete} until the number of detected paths equals the predefined number of channel paths $\widehat{L}_{\rm BU}$.

After finishing the beam training for the BS-UT link, the UT feeds back the estimated channel parameters, $\overline{\theta}_{\rm BU}^{(k)}$, $\overline{\varphi}_{\rm BU}^{(k)}$ and $\overline{r}_{\rm BU}^{(k)}$ for $k=1,2,\cdots,\widehat{L}_{\rm BU}$, to the BS so that the BS can calculate the position of the UT.
%the condition that the accumulated power of the echoes from the target is smaller than a threshold $\varrho_{\rm u}$ is satisfied. Note that $\varrho_{\rm u}$ can be determined referring to the criterion of the constant false alarm rate (CFAR) detector ~\cite{FRSP}. 

Finally, we summarize the proposed SBTTS scheme in \textbf{Algorithm}~\textbf{\ref{alg1}}, which includes the IPEBTTS algorithm from step 5 to step 18.

We also analyze the computational complexity of the proposed SBTTS scheme. From \textbf{Algorithm~\ref{alg1}}, the SBTTS scheme includes $\widehat{L}_{\rm BR}$ times of parameter estimation for beam training, $\widehat{T}$ times of parameter estimation for target sensing, and $\widehat{L}_{\rm BU}$ times of path detection for the BS-UE link. In addition, the parameter estimation for beam training includes transformation in \eqref{ADvsDD}, maximum search in~\eqref{BeamTrainingResult}, channel AoA estimation in \eqref{OneWayAoA}, channel delay estimation in \eqref{tauBR}, and channel AoD estimation in~\eqref{leastsquare2}. Since the transformation  in \eqref{ADvsDD} can be performed based on the fast Fourier transform, its computational complexity is $\mathcal{O}(N_{\rm T}N_{\rm RIS}\log_2(M))$. The computational complexity of maximum search in~\eqref{BeamTrainingResult} is $\mathcal{O}(N_{\rm T}N_{\rm RIS}M)$. Based on the closed-form expression in \eqref{GammaEquation}, the computational complexities of channel AoA estimation in \eqref{OneWayAoA} and  channel delay estimation in \eqref{tauBR} are negligible. Suppose the one-dimension search with $N_{\rm Sear}$ times of searching is used to solve \eqref{leastsquare2}. Then the computational complexity of the parameter estimation for beam training is  $\mathcal{O}(\widehat{L}_{\rm BR}N_{\rm T}N_{\rm RIS}M + \widehat{L}_{\rm BR}N_{\rm Sear})$. Similarly, the computational complexities of the parameter estimation for target sensing and the  path detection for the BS-UE link are $\mathcal{O}(\widehat{T}N_{\rm T}N_{\rm RIS}M + \widehat{T}N_{\rm Sear})$ and $\mathcal{O}(\widehat{L}_{\rm BU}N_{\rm T}N_{\rm UT}M)$, respectively. In summary, the computational complexity of the proposed SBTTS scheme is $\mathcal{O}((\widehat{T} + \widehat{L}_{\rm BR}+\widehat{L}_{\rm BU} )N_{\rm T}N_{\rm RIS}M + (\widehat{T} + \widehat{L}_{\rm BR})N_{\rm Sear})$.
\section{Positioning and Array Orientation Estimation}\label{PAO}
%In existing works, the wireless communication channels can be divided into the LoS channels and the NLoS channels, according to whether the LoS path between the transmitter and the receiver is blocked or not. For high frequency band, such as millimeter wave or terahertz, communications rely highly on the LoS path due to the large propagation attenuation of the NLoS paths. Therefore, for systems with high  carrier frequency, it would be more meaningful for beam alignment based on  LoS channels. On the other hand, for conventional sub-6G, communications can perform with both LoS channels and NLoS channels. Therefore, for systems with low  carrier frequencies, beam alignment for both LoS channels and NLoS channels is worth investigating. 

In this section, we propose the PAOE scheme for both the LoS channels and the NLoS channels based on the beam training results of SBTTS, where the TDFS algorithm is proposed to obtain the position and array orientation for the NLoS channels. 
%In this section, we assume the LoS path exists both between the BS and the RIS and between the BS and the UT. To achieve beam alignment for the RIS aided systems, we first reuse the training pilots for the BS-RIS-BS link in Sec. \ref{SBT} to perform beam training for the BS-UT link. Then, we try to utilize the beam training results in Sec.~\ref{SBT} and the training results for the BS-UT link to assist the beam alignment of the RIS-UT link.   
\subsection{For LoS Channels}\label{PositioningLoS}
To obtain the positions of  the BS, the RIS, and the UT, we first establish a Cartesian coordinate system with the position of the BS as the origin, the direction along the BS antenna array as the x-axis, and the normal direction of the BS antenna array as the y-axis. Without loss of generality, we specify the first path as the LoS path. The positions of the BS, the RIS, and the UT can be expressed as
\begin{align}\label{PositionUT}
	\setlength{\abovedisplayskip}{3pt}
	\setlength{\belowdisplayskip}{3pt}
	&\boldsymbol{p}_{\rm BS} = [0,0]^{\rm T},~\widehat{\boldsymbol{p}}_{\rm RIS} = \overline{r}_{\rm BR}^{(1)}\left[\overline{\theta}_{\rm BR}^{(1)},\sqrt{1-\big(\overline{\theta}_{\rm BR}^{(1)}\big)^2}\right]^{\rm T}\nonumber \\
	&\widehat{\boldsymbol{p}}_{\rm UT} =  \overline{r}_{\rm BU}^{(1)}\left[\overline{\theta}_{\rm BU}^{(1)},\sqrt{1-\big(\overline{\theta}_{\rm BU}^{(1)}\big)^2}\right]^{\rm T}.
\end{align}
Then array orientations of the BS and the UT can be expressed as 
\begin{align}\label{AOUT}
	\setlength{\abovedisplayskip}{3pt}
	\setlength{\belowdisplayskip}{3pt}
	&\widehat{\boldsymbol{q}}_{\rm UT} = \left[ \begin{matrix}
		\sqrt{1- \big(\overline{\varphi}_{\rm BU}^{(1)}\big)^2}	&	-\overline{\varphi}_{\rm BU}^{(1)}	\\
		\overline{\varphi}_{\rm BU}^{(1)}	&	\sqrt{1- \big(\overline{\varphi}_{\rm BU}^{(1)}\big)^2}	\\
	\end{matrix} \right] \frac{-\widehat{\boldsymbol{p}}_{\rm UT}}{\|\widehat{\boldsymbol{p}}_{\rm UT}\|_2}	\nonumber \\
	&{\boldsymbol{q}}_{\rm BS} = [0,1]^{\rm T}.
\end{align}

\begin{figure}[!t]
	\begin{center}
		\includegraphics[width=75mm]{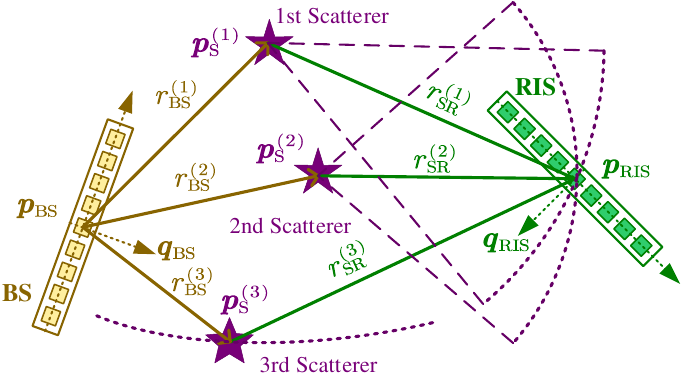}
	\end{center}
	\caption{Illustration of NLoS positioning of RIS.}
	\label{Relation2}
	\vspace{-0.8cm}
\end{figure}
%Note that for each $\widetilde{\vartheta}_1$, two estimations of $\overline{\varphi}_{\rm BR}^{(1)}$ are obtained in \eqref{OneWayAoA}. However, with only LoS path information, the ambiguity between the two estimations can not be removed in this stage. Therefore, we still leave the estimation of  $\overline{\varphi}_{\rm BR}^{(1)}$ ato the next section.
\begin{figure*}
	\begin{subequations}\label{Positioning2}
		\begin{align}
			&\min_{\boldsymbol{z}}~\sum_{u=1}^{\widehat{L}_{\rm BR}} |\overline{g}_{\rm BR}^{(u)}|^2\bigg(|\overline{\varphi}_{\rm BR}^{(u)} - \widetilde{\varphi}_{\rm BR}^{(u)}|^2 + |\overline{\theta}_{\rm BR}^{(u)} - \widetilde{\theta}_{\rm BR}^{(u)}|^2 + 4\Delta f^2|\overline{r}_{\rm BR}^{(u)} - \widetilde{r}_{\rm BR}^{(u)}|^2/c^2\bigg) \label{Objective1} \\
			&{\rm~s.t.}~~\boldsymbol{z} = [x_1,y_1,\cdots,x_{\widehat{L}_{\rm BR}},y_{\widehat{L}_{\rm BR}},x_{\rm R},y_{\rm R},\boldsymbol{q}_{\rm RIS}^{\rm T}]^{\rm T}~\label{st3}\\
			&~~~~~~~\widetilde{\theta}_{\rm BR}^{(u)}\!=\! {\rm atan}(x_u/y_u),~\widetilde{r}_{\rm BR}^{(u)}\! =\! \sqrt{x_u^2 + y_u^2} +  \sqrt{(x_u-x_{\rm R})^2+(y_u-y_{\rm R})^2},~\overline{\varphi}_{\rm BR}^{(u)} \in \overline{\boldsymbol{\varUpsilon}}_{\rm BR}^{(u)}~\label{st4}\\
			&~~~~~~~\widetilde{\varphi}_{\rm BR}^{(u)}\!=\!([\boldsymbol{q}_{\rm RIS}]_2 \!-\! [\boldsymbol{l}_u]_2\boldsymbol{l}_u^{\rm T}\boldsymbol{q}_{\rm RIS})/[\boldsymbol{l}_u]_1,\boldsymbol{l}_u\!=\!\frac{[x_u\!-\!x_{\rm R},y_u\!-\!y_{\rm R}]^T}{\sqrt{(x_u\!-\!x_{\rm R})^2\!+\!(y_u\!-\!y_{\rm R})^2}},\!\|\boldsymbol{q}_{\rm RIS}\|_2\!=\!1. \label{st5}
		\end{align}
	\end{subequations}
	\hrulefill
	% \vspace*{4pt}
	\vspace{-0.5cm}
\end{figure*}
\subsection{For NLoS Channels}\label{NLoSPositioning}
Since the positioning of the RIS is more complicated than that of the UT because of the ambiguity of the AoAs for the BS-RIS link, we take the positioning of the RIS as an example.

%To obtain the position of the RIS, we first resort to \textbf{Algorithm~1}, where we have obtained channel AoD $\overline{\theta}_{\rm BR}^{(u)}$, channel AoA $\overline{\varphi}_{\rm BR}^{(u)}$,  distance of the $u$th path $\overline{r}_{\rm BR}^{(u)}$, and the channel gain $\overline{g}_{\rm BR}^{(u)}$, by performing the SBTTS scheme. 

Denote the position of the $u$th scatterer and the position of the RIS as $\boldsymbol{p}_{\rm S}^{(u)} = [x_u,y_u]^{\rm T}$ and $\boldsymbol{p}_{\rm RIS} = [x_{\rm R},y_{\rm R}]^{\rm T}$, respectively. Then, based on the estimated channel parameters, we formulate the RIS positioning problem as a least-squares problem in \eqref{Positioning2}, which is at the top of next page. In fact, \eqref{Positioning2} aims to find the position and array orientation of the RIS best fitting for the estimated channel parameters in \textbf{Algorithm~\ref{alg1}} based on the geometry relationship in Fig.~3, where the constraints are the calculated channel parameters with fixed position of RIS and the objective is to minimize the deviation between the calculated channel parameters and the estimated channel parameters. Unfortunately, \eqref{Positioning2} is a non-convex problem. Even if a local optimal solution with a small objective value can be obtained, the positioning results may also be quite different from the true ones. Generally, an effective way to solve \eqref{Positioning2} is to traverse all the possible values for the variables in $\boldsymbol{z}$, and the one achieving the smallest objective is taken as the final result. However, the computational complexity of this method is extremely high. Specifically, if $B$ values need to be tested for each variable in $\boldsymbol{z}$, totally  $2^{\widehat{L}_{\rm BR}}B^{2(\widehat{L}_{\rm BR}+2)}$  tests are needed. For example, if $B = 100$ and $\widehat{L}_{\rm BR} = 6$,  $6.4 \times 10^{33}$  tests are needed. To reduce the complexity of exhaustive search, we propose a low-complexity TDFS algorithm by exploiting the intrinsic geometric relationship between the scatterers and the RIS.

%Note that the powers of the first two estimated paths are bigger than the powers of other paths, which results in more accurate estimation of channel parameters.
\subsubsection{Initialization}
First of all, we  select the first two paths of the $\widehat{L}_{\rm BR}$ estimated paths as the anchors to compute the possible positions of the RIS.  Then, we aim to test all the possible positions of the two scatterers on the two paths. Note that the angles of the scatterers relative to the BS array are exactly the channel AoAs. Therefore, we only need to search the positions of the scatterers along the lines determined by the estimated channel AoAs. Since the estimated distance between the BS and the RIS through the $u$th path is $\overline{r}_{\rm BR}^{(u)}$, the left boundary and the right boundary for the search of the first scatterer are set as $R_{\rm L}^{(1)} = 0$ and $R_{\rm R}^{(1)} = \overline{r}_{\rm BR}^{(1)}$, respectively. Similarly, we have $R_{\rm L}^{(2)} = 0$ and $R_{\rm R}^{(2)} = \overline{r}_{\rm BR}^{(2)}$ for the second scatterer. Moreover, we set the quantized number as $B$. We adopt an iterative search method with a maximum number of iterations being $I$ and initialize the current iteration number as $ i = 0$. 

\subsubsection{Low-complexity search}\label{search}
In the $i$th search  (for $i\ge 1$), we denote the quantization search sets of the  two scatterers as $\boldsymbol{\Psi}_1$ and $\boldsymbol{\Psi}_2$, respectively, where 
\begin{align}\label{QuantizedSearchSet}
	&[\boldsymbol{\Psi}_1]_b = R_{\rm L}^{(1)} + \frac{(b-1)\left(R_{\rm R}^{(1)} - R_{\rm L}^{(1)}\right)}{B-1},\nonumber\\
	&[\boldsymbol{\Psi}_2]_d = R_{\rm L}^{(2)} + \frac{(d-1)\left(R_{\rm R}^{(2)} - R_{\rm L}^{(2)}\right)}{B-1},
\end{align}
for $b,d = 1,\cdots,B$. When testing the $b$th element in $\boldsymbol{\Psi}_1$ and the $d$th element in $\boldsymbol{\Psi}_2$, the positions of the first two  scatterers can be expressed as 
\begin{align}\label{PositionScatterers}
	& x_1 = [\boldsymbol{\Psi}_1]_b \cdot \cos\left(\pi/2 - \mathrm{asin}\left(\overline{\theta}_{\rm BR}^{(1)}\right)\right),  \nonumber \\
	& y_1 = [\boldsymbol{\Psi}_1]_b \cdot \sin\left(\pi/2 - \mathrm{asin}\left(\overline{\theta}_{\rm BR}^{(1)}\right)\right),  \nonumber \\
	& x_2 = [\boldsymbol{\Psi}_2]_d \cdot \cos\left(\pi/2 - \mathrm{asin}\left(\overline{\theta}_{\rm BR}^{(2)}\right)\right),  \nonumber \\
	& y_2 = [\boldsymbol{\Psi}_2]_d \cdot \sin\left(\pi/2 - \mathrm{asin}\left(\overline{\theta}_{\rm BR}^{(2)}\right)\right).
\end{align}
Denote the distance from the BS to the $u$th scatterer  and the distance from the $u$th scatterer to the RIS as $r_{\rm BS}^{(u)}$ and $r_{\rm SR}^{(u)}$, respectively. Then, we have~$ r_{\rm BS}^{(1)} = [\boldsymbol{\Psi}_1]_b,~r_{\rm SR}^{(1)} = \overline{r}_{\rm BR}^{(1)} - r_{\rm BS}^{(1)},~r_{\rm BS}^{(2)} = [\boldsymbol{\Psi}_2]_b~\mbox{and}~r_{\rm SR}^{(2)} = \overline{r}_{\rm BR}^{(2)} - r_{\rm BS}^{(2)}$. As shown in Fig.~\ref{Relation2}, the position of the RIS is exactly the intersection of two circles, where the center and radius of the first circle are $[x_1,y_1]$ and $r_{\rm SR}^{(1)}$, respectively, while the center and radius of the second circle are $[x_2,y_2]$ and $r_{\rm SR}^{(2)}$, respectively. With the basic geometric knowledge, we can obtain the positions of the two intersections of the two circles. Since the computation process is simple, we omit the exact expressions of the intersections for simplicity and denote the coordinates of the two intersections as $[\widetilde{x}_{\rm R}^{(1)},\widetilde{y}_{\rm R}^{(1)}]^{\rm T}$ and $[\widetilde{x}_{\rm R}^{(2)},\widetilde{y}_{\rm R}^{(2)}]^{\rm T}$, respectively. Then we find the position of the RIS best fitting for the estimated channel parameters in this test by solving
\begin{align}\label{Positioning3}
	&\min_{x_{\rm R},y_{\rm R}}~\sum_{u=1}^{2} |\overline{g}_{\rm BR}^{(u)}|^2|\overline{\varphi}_{\rm BR}^{(u)} - \widetilde{\varphi}_{\rm BR}^{(u)}|^2 \nonumber \\
	&{\rm~s.t.}~~~[x_{\rm R},y_{\rm R}] = [\widetilde{x}_{\rm R}^{(1)},\widetilde{y}_{\rm R}^{(1)}]^{\rm T}~\mathrm{or}~[x_{\rm R},y_{R}] = [\widetilde{x}_{\rm R}^{(2)},\widetilde{y}_{\rm R}^{(2)}]^{\rm T}, \nonumber \\
	&~~~~~~~~\overline{\varphi}_{\rm BR}^{(u)} \in \overline{\boldsymbol{\varUpsilon}}_{\rm BR}^{(u)},~\mathrm{and}~\eqref{st5}
\end{align}
%where 
%\begin{align*}
%	\boldsymbol{q}_{\rm RIS} =  \left[ \begin{matrix}
%		\sqrt{1- (\overline{\varphi}_{\rm BR}^{(1)})^2}	&	-\overline{\varphi}_{\rm BR}^{(1)}	\\
%		\overline{\varphi}_{\rm BR}^{(1)}	&	\sqrt{1- (\overline{\varphi}_{\rm BR}^{(1)})^2}	\\
%	\end{matrix} \right] \boldsymbol{l}_1.
%\end{align*}
The optimal solution of \eqref{Positioning3} can be obtained via eight tests. We denote the estimated position of RIS as $[\overline{x}_{\rm R}^{(b,d)},\overline{y}_{\rm R}^{(b,d)}]$, the estimated normal direction of the RIS as $\overline{\boldsymbol{q}}_{\rm RIS}^{(b,d)}$ and the minimum value of the objective in \eqref{Positioning3} as $h(b,d)$.

\begin{algorithm}[!t]
	\caption{Positioning and Array Orientation Estimation (PAOE) Scheme}
	\label{alg3}
	\begin{algorithmic}[1]
		\STATE \textbf{Input:} $\overline{\theta}_{\rm BR}^{(u)}$, $\overline{\boldsymbol{\varUpsilon}}_{\rm BR}^{(u)}$, $\overline{r}_{\rm BR}^{(u)}$ and $\widehat{g}_{\rm BR}^{(u)}$ for $u = 1,2,\cdots \widehat{L}_{\rm BR}$.
		\STATE ~~~~~~~~~$\overline{\theta}_{\rm BU}^{(k)}$, $\overline{\varphi}_{\rm BU}^{(k)}$ and $\overline{r}_{\rm BU}^{(k)}$ for $k=1,2,\cdots,\widehat{L}_{\rm BU}$.
		\STATE \textit{/*For the LoS Channels*/}
		\STATE  Obtain $\widehat{\boldsymbol{p}}_{\rm UT}$ and $\widehat{\boldsymbol{p}}_{\rm RIS}$ via \eqref{PositionUT}. 
		\STATE Obtain $\widehat{\boldsymbol{q}}_{\rm UT}$ via \eqref{AOUT}.
		\STATE  \textbf{Output:} $\widehat{\boldsymbol{p}}_{\rm UT}$, $\widehat{\boldsymbol{q}}_{\rm UT}$ and $\widehat{\boldsymbol{p}}_{\rm RIS}$. 
		\STATE \textit{/*For the NLoS Channels*/}
		\STATE\textbf{Initialization:} $I$,~$i\leftarrow 0$, $R_{\rm L}^{(1)} \leftarrow 0$ and $R_{\rm R}^{(1)} \leftarrow \overline{r}_{\rm BR}^{(1)}$, $R_{\rm L}^{(2)} \leftarrow 0$ and $R_{\rm R}^{(2)} \leftarrow \overline{r}_{\rm BR}^{(2)}$
		\WHILE{$i\le I$}
		\STATE Obtain $\boldsymbol{\Psi}_1$ and $\boldsymbol{\Psi}_2$ via \eqref{QuantizedSearchSet}.
		\FOR{$b=1,2,\cdots B$}
		\FOR{$d=1,2,\cdots B$}
		\STATE Obtain $[\overline{x}_{\rm R}^{(b,d)},\!\overline{y}_{\rm R}^{(b,d)}]$ via \eqref{Positioning3}.
		\STATE Obtain $\overline{h}(b,d)$ via \eqref{Cost}.
		\ENDFOR
		\ENDFOR 
		\STATE Obtain $\{b^{*},d^{*}\}$ via \eqref{BestIndex}. 
		\STATE Obtain $[\overrightarrow{x}^{(i)}_{\rm R},\overrightarrow{x}^{(i)}_{\rm R}]$ via \eqref{BestPostition}.
		\STATE $i\leftarrow i+1$. 
		\STATE Update $R_{\rm L}^{(1)}$, $R_{\rm R}^{(1)}$, $R_{\rm L}^{(2)}$ and $R_{\rm R}^{(2)}$ via \eqref{Update}.
		\ENDWHILE
		\STATE $\widehat{\boldsymbol{p}}_{\rm RIS}=[\overrightarrow{x}^{(I)}_{\rm R},\overrightarrow{x}^{(I)}_{\rm R}]^T$ and $\widehat{\boldsymbol{q}}_{\rm RIS} =\widetilde{\boldsymbol{q}}_{\rm RIS}^{(I)}$.
		\STATE Obtain $\widehat{\boldsymbol{p}}_{\rm UT}$ and $\widehat{\boldsymbol{q}}_{\rm UT}$ with procedures similar to steps~8$\sim$22. 
		\STATE \textbf{Output:} $\widehat{\boldsymbol{p}}_{\rm UT}$, $\widehat{\boldsymbol{q}}_{\rm UT}$, $\widehat{\boldsymbol{p}}_{\rm RIS}$ and $\widehat{\boldsymbol{q}}_{\rm RIS}$.
	\end{algorithmic}
\end{algorithm}
Then, we determine the position of the $u$th scatterer, for $u\ge3$. As shown in Fig.~\ref{Relation2}, the summation of the distance from the $u$th scatterer to the BS and the distance from the $u$th scatterer to the RIS is a constant $\overline{r}_{\rm BR}^{(u)}$, which implies that the $u$th scatterer lies on an ellipse with focus, $[\overline{x}_{\rm R}^{(b,d)},\overline{y}_{\rm R}^{(b,d)}]$ and $[0,0]$. In addition, the $u$th scatterer also lies on a line with slope $\tan(\mathrm{asin}(\overline{\varphi}_{\rm BR}^{(u)}))$. The two observations indicate that the position of the $u$th scatterer is one of the two intersections of an ellipse and a line. With basic geometric knowledge, the two intersections can be calculated as  $[\widetilde{x}_{u}^{(1)},\widetilde{y}_{u}^{(1)}]^{\rm T}$ and $[\widetilde{x}_{u}^{(2)},\widetilde{y}_{u}^{(2)}]^{\rm T}$. Then, we find the position of the $u$th scatterer   best fitting for the estimated channel parameters by solving 
\begin{align}\label{PositioningScatter}
	&\min_{x_{u},y_{u}}~ |\overline{g}_{\rm BR}^{(u)}|^2|\overline{\varphi}_{\rm BR}^{(u)} - \widetilde{\varphi}_{\rm BR}^{(u)}|^2 \nonumber \\
	&{\rm~s.t.}~~~[x_{u},y_{u}] = [\widetilde{x}_{u}^{(1)},\widetilde{y}_{u}^{(1)}]^{\rm T}~\mathrm{or}~[x_{u},y_{u}] = [\widetilde{x}_{u}^{(2)},\widetilde{y}_{u}^{(2)}]^{\rm T}, \nonumber \\
	&~~~~~~~~\overline{\varphi}_{\rm BR}^{(u)} \in \overline{\boldsymbol{\varUpsilon}}_{\rm BR}^{(u)},\widetilde{\varphi}_{\rm BR}^{(u)}\! = \! ([\overline{\boldsymbol{q}}_{\rm RIS}^{(b,d)}]_2 - [\boldsymbol{\widetilde{l}}_u]_2\boldsymbol{\widetilde{l}}_u^{\rm T}\overline{\boldsymbol{q}}_{\rm RIS}^{(b,d)})/[\boldsymbol{\widetilde{l}}_u]_1,  \nonumber \\
	&~~~~~~~~\boldsymbol{\widetilde{l}}_u = \frac{[x_u-\overline{x}_{\rm R}^{(b,d)},
		y_u-\overline{y}_{\rm R}^{(b,d)}]^{\rm T}}{\sqrt{(x_u-\overline{x}_{\rm R}^{(b,d)})^2+(y_u-\overline{y}_{\rm R}^{(b,d)})^2}}.
\end{align}
The optimal solution of \eqref{PositioningScatter} can be obtained by four tests. Note that the positioning of the $L-2$ scatterers is independent of each other and can be performed in parallel. We denote the minimum value of the objective in \eqref{PositioningScatter} as $\widetilde{h}_u$. Then, the cost value for the $b$th test in $\boldsymbol{\Psi}_1$ and the $d$th test in $\boldsymbol{\Psi}_2$ is computed as
\begin{align}\label{Cost}
	\overline{h}(b,d) = h(b,d) + \sum_{u=3}^{\widehat{L}_{\rm BR}} \widetilde{h}_u.
\end{align}
We find the index of the test with the minimum objective value in the $i$th search  by 
\begin{align}\label{BestIndex}
	\{b^{*},d^{*}\} = \arg~\min_{b,~d\in\{1,\cdots,B\}} \overline{h}(b,d).
\end{align}
Then, the position of the RIS in the $i$th iteration best fitting for the estimated channel parameters can be  expressed as 
\begin{align}\label{BestPostition}
	[\overrightarrow{x}^{(i)}_{\rm R},\overrightarrow{x}^{(i)}_{\rm R}] = [\overline{x}^{(b^{*},d^{*})}_{\rm R},\overline{x}^{(b^{*},d^{*})}_{\rm R}]. 
\end{align}
The corresponding normal direction of the RIS is $\widetilde{\boldsymbol{q}}_{\rm RIS}^{(i)} = \overline{\boldsymbol{q}}_{\rm RIS}^{(b^*,d^*)}$. To prepare for the search of the next iteration, we update the search boundaries as 
\begin{align}\label{Update}
	&R_{\rm L}^{(1)}\leftarrow [\boldsymbol{\Psi}_1]_{b^*-1},~R_{\rm R}^{(1)}\leftarrow [\boldsymbol{\Psi}_1]_{b^*+1},\nonumber \\
	&R_{\rm L}^{(1)}\leftarrow [\boldsymbol{\Psi}_2]_{d^*-1},~R_{\rm R}^{(2)}\leftarrow [\boldsymbol{\Psi}_2]_{d^*+1}.
\end{align}
\subsubsection{Stop Condition and Outputs} 
We repeat the search procedure in Section~\ref{search} until the maximum number of iterations $I$ is achieved. The estimated position of the RIS is $[\overrightarrow{x}^{(I)}_{\rm R},\overrightarrow{x}^{(I)}_{\rm R}]$, and the estimated normal direction of the RIS is $\widetilde{\boldsymbol{q}}_{\rm RIS}^{(I)}$.
\begin{table*}[!t]%调节图片位置，h：浮动；t：顶部；b:底部；p：当前位置
	\centering
	\renewcommand{\arraystretch}{1.5}
	\caption{Comparison of schemes and overheads for different cases}
	\label{tab1}  
	\begin{tabular}{cccc cc}%表格中的数据居中，c的个数为表格的列数
		\hline\hline\noalign{\smallskip}	
		Cases &  BS-UT link  & BS-RIS link & RIS-UT link & Scheme & Overhead \\
		\noalign{\smallskip}\hline\noalign{\smallskip}
		\textbf{Case 1} & LoS & LoS & LoS & SBTTS + Section~\ref{PositioningLoS} + Section~\ref{LOSRISUT2}  &$N_{\rm BS}N_{\rm RIS}+2$  \\
		\textbf{Case 2} & LoS & LoS & NLoS &SBTTS + RIS-UT link beam training & $N_{\rm BS}N_{\rm RIS} + N_{\rm RIS} N_{\rm UT}$    \\
		\textbf{Case 3} & LoS & NLoS & LoS &SBTTS + Section~\ref{NLoSPositioning} + Section~\ref{LOSRISUT1} & $N_{\rm BS}N_{\rm RIS}$  \\
		\textbf{Case 4} & LoS & NLoS & NLoS &SBTTS + RIS-UT link beam training &$N_{\rm BS}N_{\rm RIS} + N_{\rm RIS} N_{\rm UT}$  \\
		\textbf{Case 5} & NLoS & LoS & LoS &SBTTS + Section~\ref{PositioningLoS} + Section~\ref{LOSRISUT2} & $N_{\rm BS}N_{\rm RIS}+2$ \\
		\textbf{Case 6} & NLoS & LoS & NLoS &SBTTS + RIS-UT link beam training & $N_{\rm BS}N_{\rm RIS} + N_{\rm RIS} N_{\rm UT}$\\
		\textbf{Case 7} & NLoS & NLoS & LoS &SBTTS + Section~\ref{NLoSPositioning} + Section~\ref{LOSRISUT1} &$N_{\rm BS}N_{\rm RIS}$\\
		\textbf{Case 8} & NLoS & NLoS & NLoS &SBTTS + RIS-UT link beam training & $N_{\rm BS}N_{\rm RIS} + N_{\rm RIS} N_{\rm UT}$ \\
		\noalign{\smallskip}\hline
	\end{tabular}
	\vspace{-0.5cm}
\end{table*}

%Note that \textbf{Algorithm~\ref{alg3}} can also be applied to positioning of the UT based on the beam training results in \textbf{Algorithm~\ref{alg1}}. 
Finally, we summarize the PAOE scheme  in  \textbf{Algorithm~\ref{alg3}}, which includes the TDFS algorithm from step 9 to step 21. 

%In addition, different from the cases of LoS  channels, the ambiguity of AoA for the BS-RIS link can be removed with the cooperation of multiple NLoS paths and the normal direction of the RIS can be determined. 

Let us discuss the computational complexity of the proposed low-complexity  TDFS algorithm. The proposed  TDFS algorithm consists of $I$ iterations. For each iteration, $B^2$ times of search are needed. For each search, $4\widehat{L}_{\rm BR}$ tests are needed. Overall, the computational complexity of the proposed positioning algorithm  is $\mathcal{O}(4IB^2\widehat{L}_{\rm BR})$. For example, if $B = 100$, $I = 5$, and $\widehat{L}_{\rm BR} = 6$, the proposed method reduces the number of tests from  $6.4\times 10^{33}$ for exhaustive search to only $1.2 \times 10^{6}$, which is a significant improvement.
\section{Beam Alignment based on SBTTS and PAOE}\label{SBA}
In this section, we investigate the beam alignment in the RIS-assisted ISAC systems based on the SBTTS and PAOE schemes. As shown in Table~\ref{tab1}, we divide the RIS-assisted ISAC systems into eight cases, which can be regarded as three kinds. The first kind contains Cases 3 and 7 with the NLoS-dominant BS-RIS links and LoS-dominant RIS-UT links. The second kind contains Cases 1 and 5 with the  LoS-dominant BS-RIS links and the LoS-dominant RIS-UT links. The third kind contains  Cases 2,  4, 6 and 8 with the NLoS-dominant RIS-UT links. Then, beam alignment for the three kinds of RIS-assisted ISAC systems is discussed.

\subsection{NLoS-dominant BS-RIS Links and LoS-dominant RIS-UT Links}\label{LOSRISUT1}

\begin{figure}[!t]
	\centering
	\includegraphics[width=60mm]{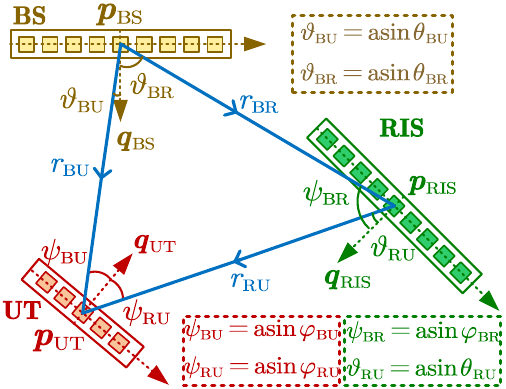}
	\caption{The geometry relationship among the BS, the RIS and the UT.}
	\label{figBSUTRelation}
	\vspace{-0.5cm}
\end{figure}

In Fig. \ref{figBSUTRelation}, we illustrate the geometry relationship among the antenna arrays at the BS, the RIS, and the UT. Note that we have obtained positions and array orientations of the RIS and UT in Section~\ref{NLoSPositioning}. Therefore, we only need to obtain $\widehat{\theta}_{\rm RU}$ and $\widehat{\varphi}_{\rm RU}$ to establish the link between the RIS and the UT. Denote $\boldsymbol{k} \triangleq (\widehat{\boldsymbol{p}}_{\rm RIS} - \widehat{\boldsymbol{p}}_{\rm UT})/\|\widehat{\boldsymbol{p}}_{\rm RIS} - \widehat{\boldsymbol{p}}_{\rm UT}\|_2$. Then,  $\widehat{\theta}_{\rm RU}$ and $\widehat{\varphi}_{\rm RU}$ can be expressed as
\begin{align}\label{ThetaPhi}
	&\widehat{\theta}_{\rm RU} = -([{\widehat{\boldsymbol{q}}}_{\rm RIS}]_2 - [\boldsymbol{k}]_2\boldsymbol{k}^{\rm T}{\widehat{\boldsymbol{q}}}_{\rm RIS})/[\boldsymbol{k}]_1, \nonumber\\
	&\widehat{\varphi}_{\rm RU} = ([\widehat{\boldsymbol{q}}_{\rm UT}]_2 - [\boldsymbol{k}]_2\boldsymbol{k}^{\rm T}\widehat{\boldsymbol{q}}_{\rm UT})/[\boldsymbol{k}]_1.
\end{align}
\subsection{LoS-dominant BS-RIS Links and LoS-dominant RIS-UT Links}\label{LOSRISUT2}
For this kind of RIS-assisted ISAC systems, the RIS-UT link can also be established similar to Section~\ref{LOSRISUT1}. Then, we have
\begin{align}\label{ThetaPhi2}
	&\widehat{\varphi}_{\rm RU} = ([\widehat{\boldsymbol{q}}_{\rm UT}]_2 - [\boldsymbol{k}]_2\boldsymbol{k}^{\rm T}\widehat{\boldsymbol{q}}_{\rm UT})/[\boldsymbol{k}]_1.
\end{align}
However, the estimation of $\widehat{\theta}_{\rm RU}$ depends on the array orientation determination of the RIS.  Recall that for each estimated $\widetilde{\vartheta}_q$ in \eqref{TwoWayAoA}, the channel AoA between the BS and RIS $\widehat{\varphi}_{\rm BR}^{(q)}$ has two possible values in \eqref{OneWayAoA}. We denote them as $\overrightarrow{\varphi}_{\rm BR}^{(1)}$ and $\overrightarrow{\varphi}_{\rm BR}^{(2)}$, respectively. Similar to \eqref{AOUT}, we can obtain the two possible array orientations of the RIS, which are denoted as $\overrightarrow{\boldsymbol{q}}_{\rm RIS}^{(1)}$ and $\overrightarrow{\boldsymbol{q}}_{\rm RIS}^{(2)}$, respectively. 
% \begin{align}\label{AORIS}
% 	&\overrightarrow{\boldsymbol{q}}_{\rm RIS}^{(1)} = \left[ \begin{matrix}
% 		\sqrt{1- \big(\overrightarrow{\varphi}_{\rm BR}^{(1)}\big)^2}	&	-\overrightarrow{\varphi}_{\rm BR}^{(1)}	\\
% 		\overrightarrow{\varphi}_{\rm BR}^{(1)}	&	\sqrt{1- \big(\overrightarrow{\varphi}_{\rm BR}^{(1)}\big)^2}	\\
% 	\end{matrix} \right] \frac{\widehat{\boldsymbol{p}}_{\rm RIS}}{\|\widehat{\boldsymbol{p}}_{\rm RIS}\|_2}	\nonumber \\
% 		&\overrightarrow{\boldsymbol{q}}_{\rm RIS}^{(2)} = \left[ \begin{matrix}
% 		\sqrt{1- \big(\overrightarrow{\varphi}_{\rm BR}^{(2)}\big)^2}	&	-\overrightarrow{\varphi}_{\rm BR}^{(2)}	\\
% 		\overrightarrow{\varphi}_{\rm BR}^{(2)}	&	\sqrt{1- \big(\overrightarrow{\varphi}_{\rm BR}^{(2)}\big)^2}	\\
% 	\end{matrix} \right] \frac{\widehat{\boldsymbol{p}}_{\rm RIS}}{\|\widehat{\boldsymbol{p}}_{\rm RIS}\|_2}.
% \end{align}
Then, two possible channel AoDs for the RIS-UT link can be expressed as 
\begin{align}\label{Theta2}
	&\overrightarrow{\theta}_{\rm RU}^{(1)} = -([\overrightarrow{\boldsymbol{q}}_{\rm RIS}^{(1)}]_2 - [\boldsymbol{k}]_2\boldsymbol{k}^{\rm T}\overrightarrow{\boldsymbol{q}}_{\rm RIS}^{(1)})/[\boldsymbol{k}]_1, \nonumber\\
	&\overrightarrow{\theta}_{\rm RU}^{(2)} = -([\overrightarrow{\boldsymbol{q}}_{\rm RIS}^{(2)}]_2 - [\boldsymbol{k}]_2\boldsymbol{k}^{\rm T}\overrightarrow{\boldsymbol{q}}_{\rm RIS}^{(2)})/[\boldsymbol{k}]_1.
\end{align}
To determine whether $\overrightarrow{\varphi}_{\rm BR}^{(1)}$  or $\overrightarrow{\varphi}_{\rm BR}^{(2)}$ is the correct channel AoA for the BS-RIS link, the UT transmits two uplink training pilots with $\boldsymbol{\alpha}(N_{\rm UT},\widehat{\varphi}_{\rm RU})$. Then, the RIS reflects the signal from the UT with $\sqrt{N_{\rm RIS}}\boldsymbol{\alpha}(N_{\rm RIS},-\overrightarrow{\varphi}_{\rm BR}^{(1)} - \overrightarrow{\theta}_{\rm RU}^{(1)})$ and $\sqrt{N_{\rm RIS}}\boldsymbol{\alpha}(N_{\rm RIS},-\overrightarrow{\varphi}_{\rm BR}^{(2)} - \overrightarrow{\theta}_{\rm RU}^{(2)})$, sequentially. After that, the BS receives the uplink pilots with $\boldsymbol{\alpha}(N_{\rm T},\overline{\theta}_{\rm BR}^{(1)})$. The received signals of the two pilots at the BS are denoted as $\overrightarrow{y}_1$ and $\overrightarrow{y}_2$, respectively. If $|\overrightarrow{y}_1|\geq |\overrightarrow{y}_2|$, we have  $\overline{\varphi}_{\rm BR}^{(1)} = \overrightarrow{\varphi}_{\rm BR}^{(1)},~\overline{\theta}_{\rm RU}^{(1)} = \overrightarrow{\theta}_{\rm RU}^{(1)},~\mbox{and}~ \widehat{\boldsymbol{q}}_{\rm RIS} = \overrightarrow{\boldsymbol{q}}_{\rm RIS}^{(1)}$, otherwise,  $\overline{\varphi}_{\rm BR}^{(1)} = \overrightarrow{\varphi}_{\rm BR}^{(2)},~\overline{\theta}_{\rm RU}^{(1)} = \overrightarrow{\theta}_{\rm RU}^{(2)},~\mbox{and}~ \widehat{\boldsymbol{q}}_{\rm RIS} = \overrightarrow{\boldsymbol{q}}_{\rm RIS}^{(2)}$.
%\begin{algorithm}[!t]
%	\caption{Sensing-assisted Beam Alignment}
%	\label{alg33}
%	\begin{algorithmic}[1]
%		\STATE \textbf{Input:} $\overline{\theta}_{\rm BR}^{(1)}$, $\overline{\varphi}_{\rm BR}^{(1)}$ and $\overline{r}_{\rm BR}^{(1)}$.
%		\STATE \textit{/*  Beam training for the BS-UT link */}
%		\STATE Obtain $\rho$ via \eqref{ratio1} and \eqref{ratio2}. 
%		\IF{$\rho \ge 0.5$}
%		\STATE Obtain ${n}_{\rm UT}^{*}$, ${t}_{\rm UT}^{*}$ and ${m}_{\rm UT}^*$ via \eqref{BeamTrainingResultUT}.
%		\ELSIF{$\rho < 0.5$}
%		\STATE Obtain ${n}_{\rm UT}^{*}$, ${t}_{\rm UT}^{*}$ and ${m}_{\rm UT}^*$ via \eqref{BeamTrainingResultUT2}.
%		\ENDIF
%		\STATE Obtain $\widehat{\theta}_{\rm BU}$ with ${n}_{\rm UT}^{*}$, $N_{\rm T}$, $\widetilde{\boldsymbol{Z}}_n$ and  \textbf{Lemma~1}.
%		\STATE Obtain $\widehat{\varphi}_{\rm BU}$ with ${t}_{\rm UT}^{*}$, $N_{\rm UT}$, $\widetilde{\boldsymbol{Z}}_n$ and  \textbf{Lemma~1}.
%		\STATE Obtain $\widehat{\theta}_{\rm BU}$ with ${m}_{\rm UT}^{*}$, $M$, $\widetilde{\boldsymbol{Z}}_n$ and  \textbf{Lemma~1}.
%		\STATE \textit{/* Beam alignment for the RIS-UT link */}
%		\STATE Obtain $\widehat{\theta}_{\rm RU}$,~$\widetilde{\varphi}_{\rm BR}^{(1)}$ and~$\widetilde{\varphi}_{\rm BR}^{(2)}$  via \eqref{ThetaPhi}.
%		\STATE \textit{/* Uplink confirmation */}
%		\STATE  Obtain $\widehat{\varphi}_{\rm BR}$ and  $\widehat{\theta}_{\rm BU}$ via \eqref{Confirmation}.
%		\STATE \textbf{Output:} $\widehat{\varphi}_{\rm BR}$, $\widehat{\varphi}_{\rm RU}$ and $\widehat{\varphi}_{\rm RU}$.
%	\end{algorithmic}
%\end{algorithm}

\subsection{Discussion}
\textbf{\textit{NLoS-dominant RIS-UT Links:}} For the case of NLoS-dominant RIS-UT link, the geometry relationship in Fig.~\ref{figBSUTRelation} for the  LoS-dominant RIS-UT link does not hold anymore. In this condition, a straightforward method is to perform additional beam training for the RIS-UT link. Note that channel AoD for the BS-RIS link has been obtained based on the SBTTS scheme, which simplifies the beam alignment of the three-node BS-RIS-UT link to the beam alignment of the two-node RIS-UT link. Therefore, only $N_{\rm RIS}N_{\rm UT}$ more times of beam training are needed.

\textbf{\textit{Training Overhead:}} In Table~\ref{tab1}, we compare the training overheads of different cases. The proposed SBTTS scheme in Section~\ref{SBT} needs $N_{\rm T}N_{\rm RIS}$ times of beam training. As the beam training for the BS-UT link reuses the training pilots for the BS-RIS link, no more training pilot is needed. If the BS-RIS link is LoS-dominant, two more training pilots are needed to resolve  the ambiguity in \eqref{OneWayAoA}. In addition, if the RIS-UT link is NLoS-dominant, additional $N_{\rm RIS}N_{\rm UT}$ times of beam training are needed to establish the RIS-UT link. In summary, for Cases 1 and 5,  $N_{\rm T}N_{\rm RIS} + 2$ times of beam training are needed to establish links among BS, RIS, and UT.  For Cases 2, 4, 6, and 8,  $N_{\rm T}N_{\rm RIS} + N_{\rm RIS}N_{\rm UT}$ times of beam training are needed to establish links among BS, RIS, and UT. For Cases 3 and 7,  $N_{\rm T}N_{\rm RIS}$ times of beam training are needed to establish links among BS, RIS, and UT. 

\textbf{\textit{Beam Tracking:}} The proposed scheme can also be combined with beam tracking techniques to improve the efficiency of beam alignment. In practical systems, the BS-RIS link, the BS-UT link, and the RIS-UT link all have the potential to be blocked. If beam tracking is integrated into the proposed scheme, we can quickly establish the BS-RIS-UT link if the BS-UT link is blocked thanks to the accurate positioning of the RIS. However, the conventional beam tracking schemes can only establish the BS-UT link based on beam tracking.

\section{Simulation Results}\label{SimulationResults}
Now we evaluate the performance of the proposed  schemes. We consider an ISAC BS employing a transceiver with $N_{\rm T} = 64$ antennas and a receiver with $N_{\rm R} = 16$ antennas. The number of elements at the RIS is set to be $128$.  OFDM is adopted for wideband processing, where the number of subcarriers is set to be $M = 128$ and the subcarrier spacing is set to be $\Delta f = 120$ kHz. To compare different systems fairly, we adopt the widely-used free space propagation model to characterize the path loss \cite{JSAC22SXD}, where $r_{\rm Tar}^{(l)}\sim U(50,150)$~meters. The noise power  at both the receiver and the UT is set to be $-103$~dBm. The number of targets is set to be $T=6$. In addition, the target velocity distributes randomly within $[10,30]$ meters per second.
\subsection{For the Case of LoS Channels}
We first evaluate the proposed schemes for Case 1, where the LoS path exists in the BS-UT link, the BS-RIS link, and the RIS-UT link. We set the carrier frequency to be $26.5$~GHz. For the channels, we set $L_{\rm BR} = 3$, $L_{\rm BU}= 3$, where $r_{\rm BR}^{(l)} \sim U(20,40)$~meters  and $r_{\rm BU}^{(l)}\sim U(50,150)$~meters. The ratio of the LoS path power to the NLoS path power is set to be 20~dB. The radar cross section (RCS) of targets distributes randomly  within $[-15,10]$ dBsm.

%\begin{figure}[!t]
%	\centering
%	\begin{minipage}[t]{0.45\textwidth}
%		\centering
%		\includegraphics[width=1\textwidth]{PE.eps}
%		\caption{Comparisons of the positioning error of different equipment for Case~1 in Table~\ref{tab1}.}
%		\label{PA}
%	\end{minipage}
%	\begin{minipage}[t]{0.45\textwidth}
%		\centering
%		\includegraphics[width=1\textwidth]{BG.eps}
%		\caption{Comparisons of the beamforming gains of different schemes for Case~1 in Table~\ref{tab1}.}
%		\label{BG}
%	\end{minipage}
%	\vspace{-1cm}
%\end{figure}

 \begin{figure}[!t]
	\begin{center}
		\includegraphics[width=70mm]{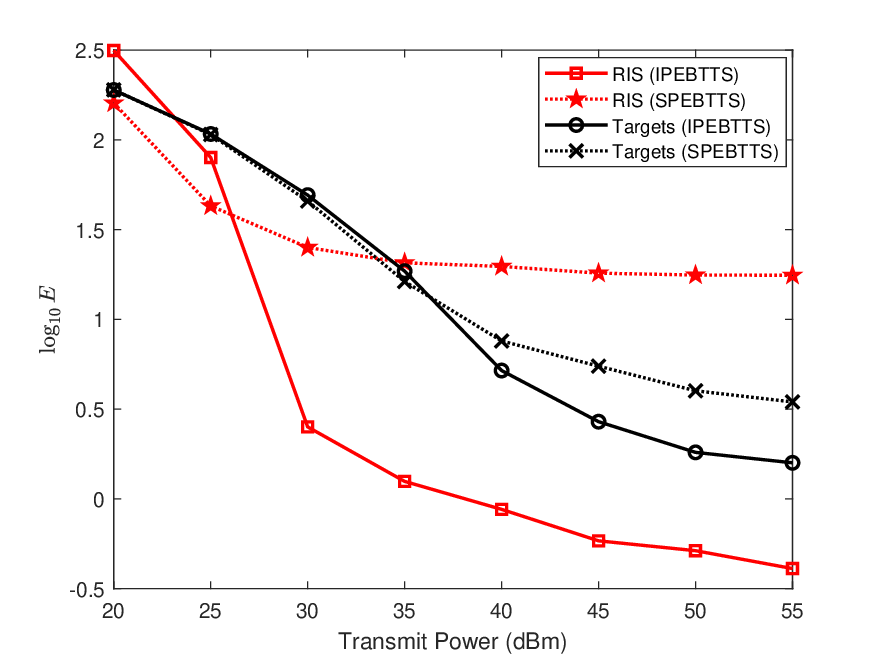}
	\end{center}
	\caption{Comparisons of the positioning error of different equipment for Case~1 in Table.~\ref{tab1}.}
	\label{PA}
		\vspace{-0.4cm}
\end{figure}

\begin{figure}[!t]
	\begin{center}
		\includegraphics[width=70mm]{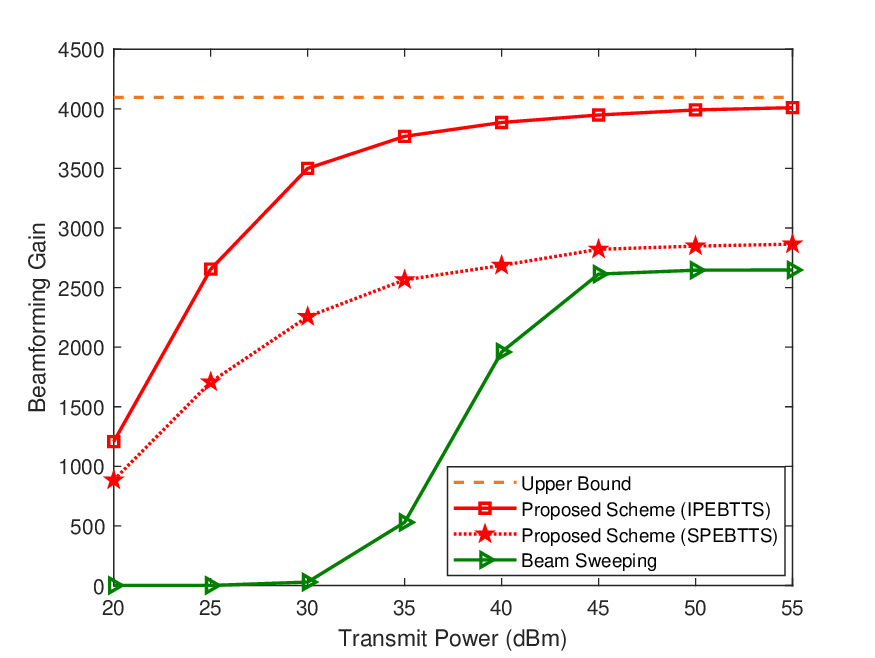}
	\end{center}
	\caption{Comparisons of the beamforming gains of different schemes for Case~1 in Table.~\ref{tab1}.}
	\label{BG}
		\vspace{-0.4cm}
\end{figure}
In Fig.~\ref{PA}, we illustrate the positioning error of the targets and the RIS. To demonstrate the effectiveness of the proposed IPEBTTS algorithm, the positioning error of the targets and the RIS with separate parameter estimation for beam training and target sensing (SPEBTTS) is adopted as the benchmark. Define the real position of the target or the RIS and its estimation as $\boldsymbol{p}$ and $\widehat{\boldsymbol{p}}$, respectively. Then, the positioning error is defined as $E \triangleq \|\boldsymbol{p}-\widehat{\boldsymbol{p}}\|_2$. In Fig.~\ref{PA}, the average positioning error of all the targets is given.  From the figure, the RIS with SPEBTTS performs better than the targets with SPEBTTS for low transmit powers. This is because the echoes from targets with small RCS are highly interfered by those from the RIS and the noise, which greatly deteriorates the positioning performance of the targets. On the contrary, the RIS with SPEBTTS performs worse than targets with SPEBTTS for high transmit powers because there are more targets than the RIS paths and the detection of the RIS is easier to be interfered than that of the targets. It is shown that the positioning performance of both the RIS and the targets is improved with the proposed IPEBTTS algorithm for high transmit power, i.e., transmit powers higher than 30~dBm. Moreover, the positioning performance of the RIS with SPEBTTS is better than that with the IPEBTTS algorithm for low transmit powers, e.g., 20 dBm. That is because SPEBTTS is performed with the interference of echoes from targets while the IPEBTTS algorithm is performed with noise, where the former leads to a smaller positioning error than the latter for low transmit powers. In addition, the RIS has better positioning performance than the targets for large transmit powers. That is because the RIS is closer to the BS than the targets in the considered scenario, which leads to the smaller positioning error of the RIS for a fixed angle error.

\begin{figure}[!t]
	\begin{center}
		\includegraphics[width=70mm]{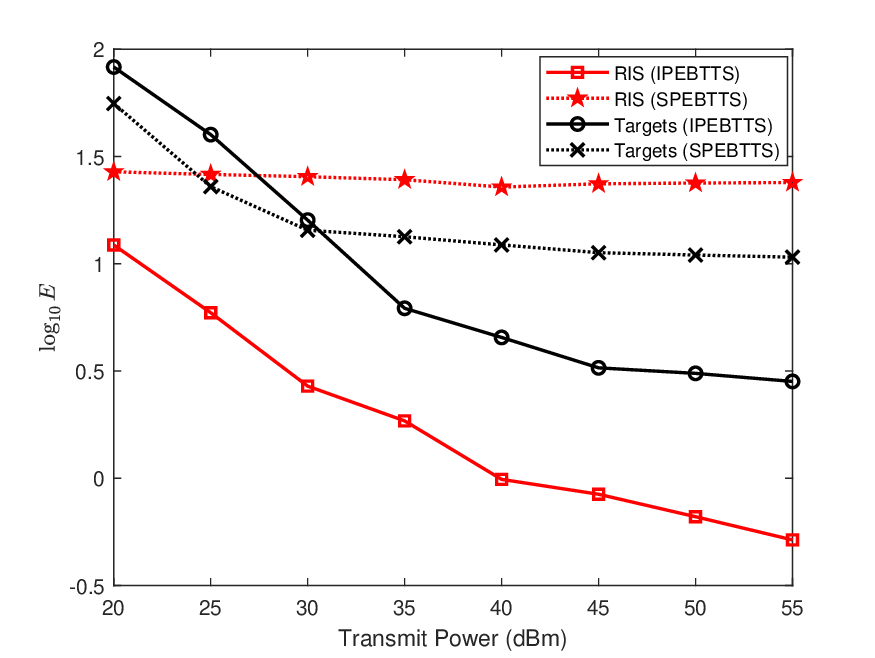}
	\end{center}
	\caption{Comparisons of the positioning error of different equipment  for Case~7 in Table.~\ref{tab1}.}
	\label{PA2}
	\vspace{-0.2cm}
\end{figure}

\begin{figure}[!t]
	\begin{center}
		\includegraphics[width=70mm]{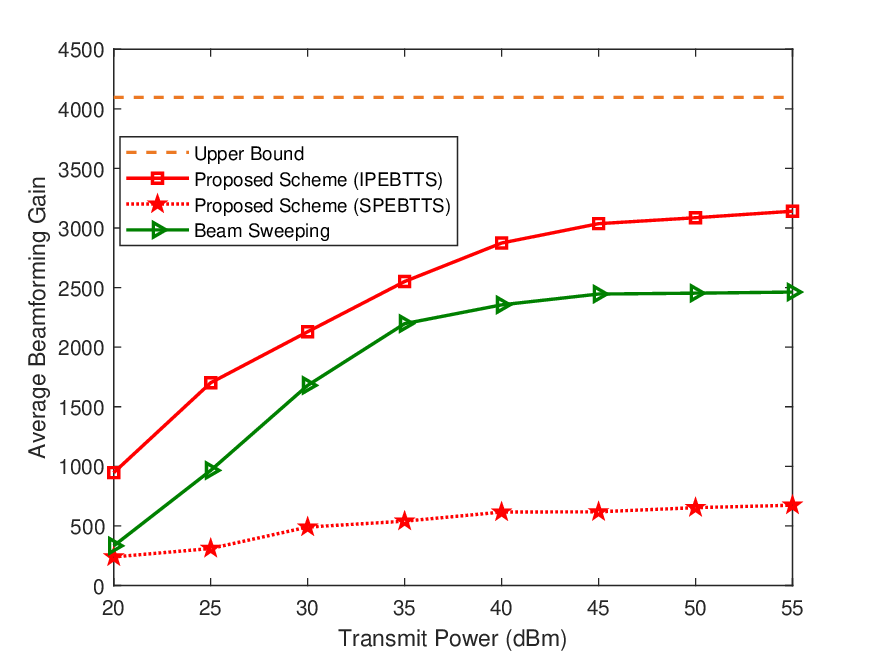}
	\end{center}
	\caption{Comparisons of the beamforming gains of different schemes for Case~7 in Table.~\ref{tab1}.}
	\label{BG2}
	\vspace{-0.5cm}
\end{figure}

In Fig.~\ref{BG}, we compare the proposed scheme with the conventional beam sweeping in terms of the beamforming gains of the BS-RIS-UT link after beam alignment for Case~1. Note that the upper bound of the beamforming gain is reached when the beam alignment at the BS, RIS and UT is accomplished simultaneously, leading to the beamforming gain being $\sqrt{N_{T}N_{\rm UT}}N_{\rm RIS} = 4096$. From the figure, the proposed scheme performs much better than the beam sweeping. For low transmit powers, such as transmit powers lower than 40~dBm, the superiority of the proposed scheme comes from the shorter distance between the BS and the RIS, which results in larger received signal-to-noise-ratios (SNRs). For high transmit powers, such as transmit powers higher than 40~dBm, the superiority of the proposed scheme comes from the accurate positioning of the RIS and the UT, which results in more accurate beam alignment than rough beam sweeping. Note that the proposed scheme needs $N_{\rm T}N_{\rm RIS}$ times of beam training while beam sweeping needs $N_{\rm T}N_{\rm RIS}N_{\rm UT}$ times of beam training, where the former is much smaller than the latter especially when  $N_{\rm UT}$ is large.

%\begin{figure}[!t]
%	\centering
%	\begin{minipage}[t]{0.45\textwidth}
%		\centering
%		\includegraphics[width=1\textwidth]{PE2.eps}
%		\caption{Comparisons of the positioning error of different equipment  for Case~7 in Table~\ref{tab1}.}
%		\label{PA2}
%	\end{minipage}
%	\begin{minipage}[t]{0.45\textwidth}
%		\centering
%		\includegraphics[width=1\textwidth]{BG.eps}
%		\caption{Comparisons of the beamforming gains of different schemes for Case~7 in Table~\ref{tab1}.}
%		\label{BG2}
%	\end{minipage}
%	\vspace{-1cm}
%\end{figure}

\subsection{For the Cases of NLoS Channels}
We  also evaluate the proposed schemes for all the other seven cases in Table~\ref{tab1} except Case 1, where we set the carrier frequency to be $5$~GHz. For the channels, we set the number of NLoS paths to six. The distance between the BS and the RIS is set to be uniformly and randomly distributed within $[20,40]$~meters  while the distance between the BS and the UT is set to be uniformly and randomly distributed within $[50,150]$~meters. The scatterers are randomly distributed in the space and the parameters of the NLoS paths are generated with the geometry relationship in Fig.~ \ref{Relation2}. The RCS of targets distributes randomly  within $[-20,10]$ dBsm.

\begin{figure}[!t]
	\begin{center}
		\includegraphics[width=70mm]{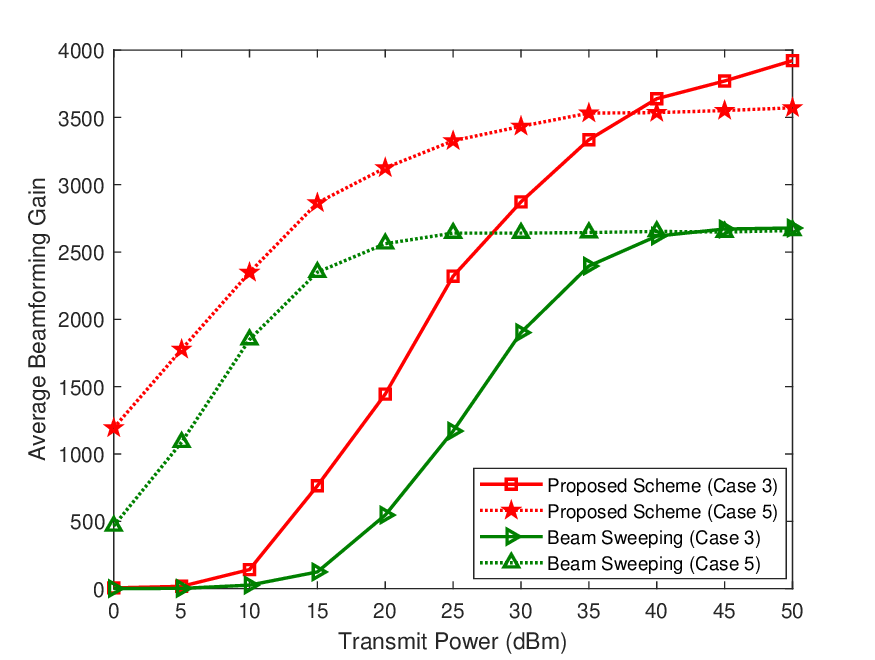}
	\end{center}
	\caption{Comparisons of the beamforming gains of different schemes  for Cases 3 and 5 in Table.~\ref{tab1}.}
	\label{BG3}
	\vspace{-0.3cm}
\end{figure}

\begin{figure}[!t]
	\begin{center}
		\includegraphics[width=70mm]{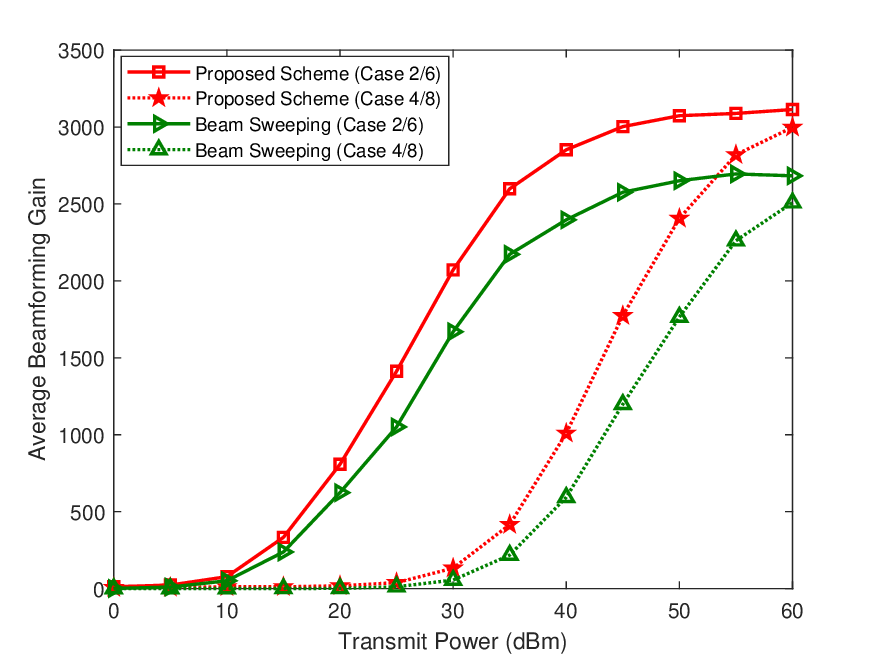}
	\end{center}
	\caption{Comparisons of the beamforming gains of different schemes  for Cases 2/6 and Cases 4/8 in Table.~\ref{tab1}.}
	\label{BG4}
	\vspace{-0.5cm}
\end{figure}

In Fig.~\ref{PA2}, we evaluate the performance of the proposed IPEBTTS and TDFS algorithms, where the positioning error of the targets and the RIS for Case 7 is adopted as the metric. It is shown that the positioning performance of  the RIS with IPEBTTS is better than that with SPEBTTS thanks to the interference cancellation between the echoes from the RIS and the echoes from the targets. It is also shown that the positioning performance of the targets with IPEBTTS is better than that with SPEBTTS for high transmit powers, such as transmit powers higher than 30~dBm, but worse than that with SPEBTTS for low transmit powers, such as transmit powers lower than 30~dBm. That is because positioning of target with SPEBTTS is performed with the interference of echoes from  the RIS while the IPEBTTS algorithm is performed with the interference of noise, where the former leads to a smaller positioning error than the latter for low transmit powers. In addition, the  positioning error of the RIS with SPEBTTS is high and does not decrease apparently with the increase of the transmit power. That is because the interference of the targets leads to the detection error of the NLoS paths and the TDFS algorithm cannot locate the RIS effectively in this condition.

In Fig.~\ref{BG2}, we compare the proposed scheme with the conventional beam sweeping in terms of the beamforming gains of the BS-RIS-UT link after beam alignment for Case 7 in Table~\ref{tab1}. Note that the average beamforming gain of all NLoS paths is taken as the metric. The low channel gains of the NLoS paths lead to the gap between the performance of the simulated schemes and the upper bound. From the figure, the proposed scheme performs better than the beam sweeping, where the superiority of the proposed scheme comes from the accurate positioning of the RIS and the UT, which results in more accurate beam alignment than rough beam sweeping.

In Fig.~\ref{BG3} and Fig.~\ref{BG4}, we compare the proposed scheme with the conventional beam sweeping in terms of the beamforming gains of the BS-RIS-UT link after beam alignment, where Cases 3 and 5 are simulated in Fig.~\ref{BG3} while Cases 2, 4, 6 and 8 are simulated in Fig.~\ref{BG4}. Note that Cases 2 and 6 have the same BS-RIS link and RIS-UT link. Therefore, we can perform simulations for Cases 2 and 6 together. The same goes for Cases 4 and 8. From the figure, the proposed scheme performs better than beam sweeping for all the cases thanks to the integration of the sensing units.

\section{Conclusion}\label{Conclusion}
In this paper, an SBTTS scheme has first been proposed, which enables the BS to perform beam training with the UTs as well as the RIS, and simultaneously to sense the targets. Then, a PAOE scheme for both the LoS channels and the NLoS channels based on the beam training results of SBTTS has been proposed, where a low-complexity TDFS algorithm has been proposed to obtain the position and array orientation for the NLoS channels. Based on the SBTTS and PAOE schemes, the AoA and AoD for the channels between the RIS and the UTs have been proposed by exploiting the geometry relationship to accomplish the beam alignment of the ISAC system. Simulation results have shown that the proposed beam alignment schemes can achieve better performance than beam sweeping with lower training overhead benefiting from the integration of sensing units. In our future work, we will try to find a proper performance trade-off between the beamforming gain and the parameter estimation error.

\bibliographystyle{IEEEtran}
\bibliography{IEEEabrv,IEEEexample}

\begin{IEEEbiography}[{\includegraphics[width=1in,height=1.25in,clip,keepaspectratio]{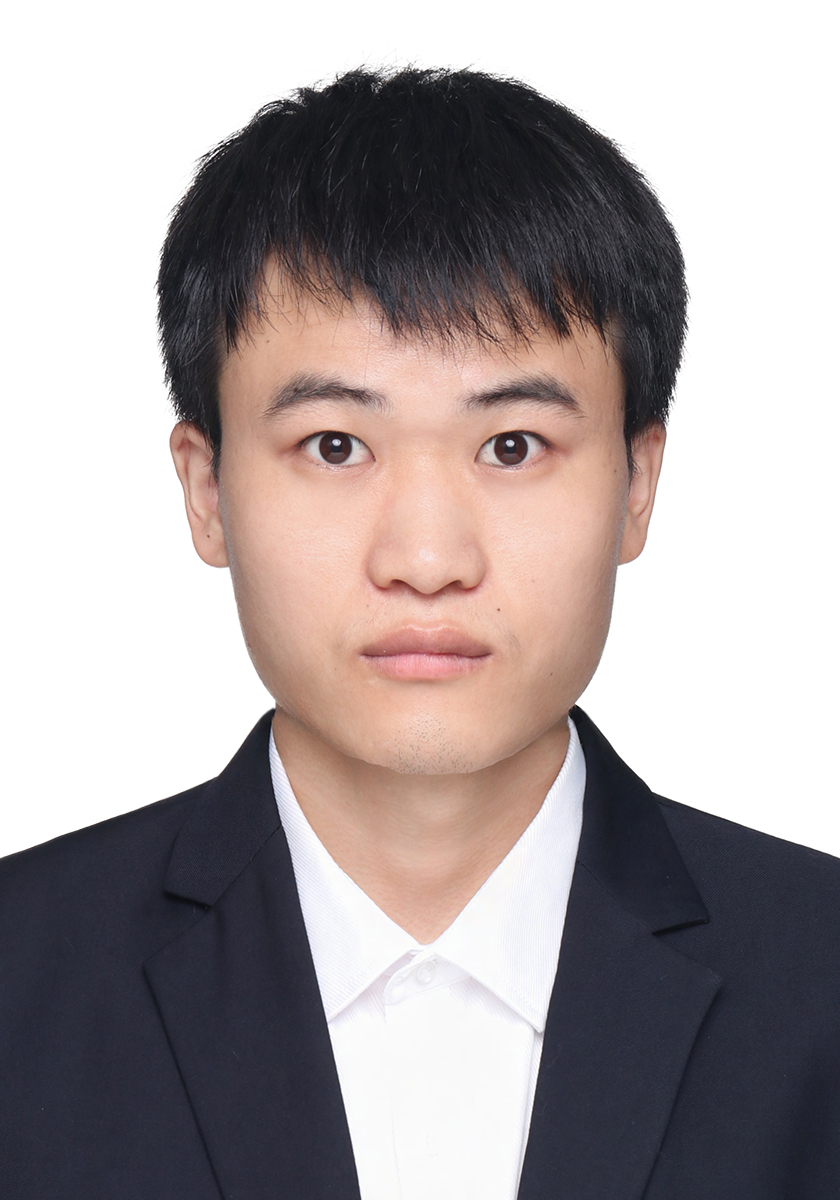}}]{Kangjian Chen}
	(Student Member, IEEE) received the B.S. degree from the Nanjing University of Science and Technology, Nanjing, China, in 2017, and the M.S. degree in signal processing from Southeast University, Nanjing, China, in 2020. He is currently pursuing the Ph.D. degree in Southeast University. His research interests include integrated sensing and communication (ISAC), massive MIMO, and reconfigurable intelligent surface (RIS). He received the Best Paper Awards from IEEE Global Communications Conference (GLOBECOM) in 2019 and IEEE/CIC International Conference on Communications in China (ICCC) in 2022.
\end{IEEEbiography}

\begin{IEEEbiography}[{\includegraphics[width=1in,height=1.25in,clip,keepaspectratio]{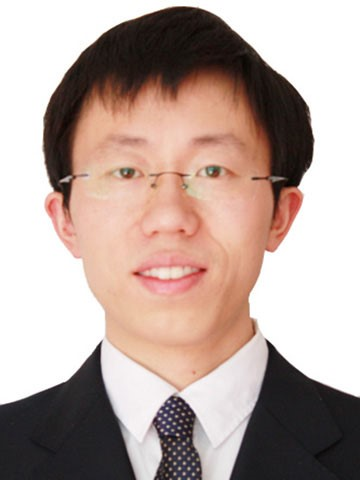}}]{Chenhao Qi}
	(Senior Member, IEEE) received the B.S. degree (Hons.) in information engineering from the Chien-Shiung Wu Honored College, Southeast University, China, in 2004, and the Ph.D. degree in signal and information processing from Southeast University in 2010. 
	
	From 2008 to 2010, he visited the Department of Electrical Engineering, Columbia University, New York, USA. Since 2010, he has been a Faculty Member with the School of Information Science and Engineering, Southeast University, where he is currently a Professor and the Head of Jiangsu Multimedia Communication and Sensing Technology Research Center. His research interests include millimeter wave communications, integrated sensing and communication (ISAC) and satellite communications. He received Best Paper Awards from IEEE GLOBECOM in 2019, IEEE/CIC ICCC in 2022, and the 11th International Conference on Wireless Communications and Signal Processing (WCSP) in 2019. He has served as an Associate Editor for IEEE TRANSACTIONS ON COMMUNICATIONS, IEEE COMMUNICATIONS LETTERS, IEEE OPEN JOURNAL OF THE COMMUNICATIONS SOCIETY, IEEE OPEN JOURNAL OF VEHICULAR TECHNOLOGY, and CHINA COMMUNICATIONS. 
\end{IEEEbiography}

\begin{IEEEbiography}[{\includegraphics[width=1in,height=1.25in,clip,keepaspectratio]{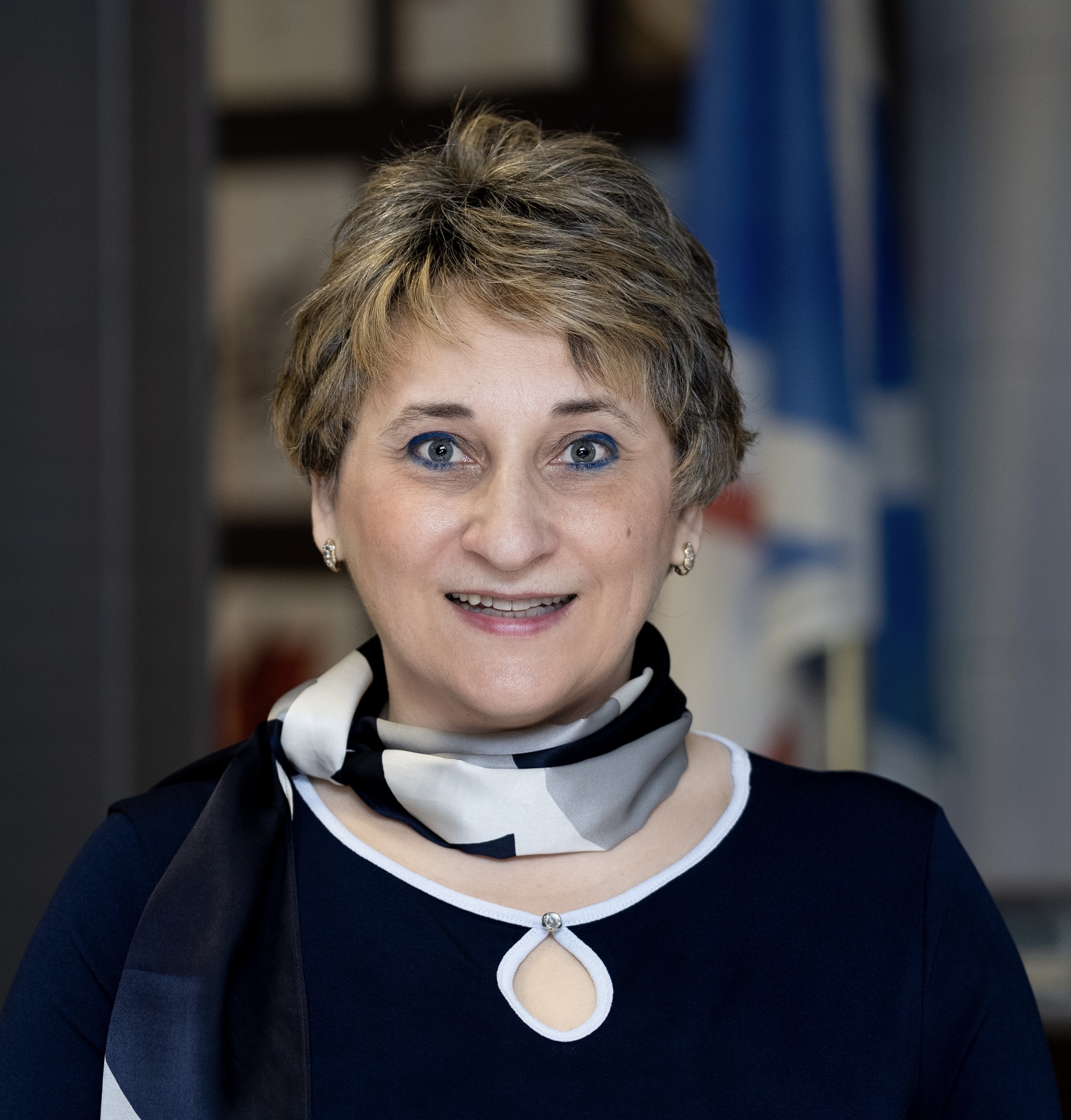}}]{Octavia A. Dobre}
	(Fellow, IEEE) is a Professor and Canada Research Chair Tier 1 at Memorial University in Canada. She was a Visiting Professor with Massachusetts Institute of Technology, USA and Universite de Bretagne Occidentale, France.
	
	Her research interests encompass wireless communication and networking technologies, as well as optical and underwater communications. She has (co-)authored over 450 refereed papers in these areas.
	
	Dr. Dobre serves as the Director of Journals of the Communications Society. She was the inaugural Editor-in-Chief (EiC) of the IEEE Open Journal of the Communications Society and the EiC of the IEEE Communications Letters.
	
	Dr. Dobre was a  Fulbright Scholar, Royal Society Scholar, and Distinguished Lecturer of the IEEE Communications Society. She obtained Best Paper Awards at various conferences, including IEEE ICC, IEEE GLOBECOM, IEEE WCNC, and IEEE PIMRC. Dr. Dobre is an elected member of the European Academy of Sciences and Arts, a Fellow of the Engineering Institute of Canada, and a Fellow of the Canadian Academy of Engineering. 
\end{IEEEbiography}

\begin{IEEEbiography}[{\includegraphics[width=1in,height=1.25in,clip,keepaspectratio]{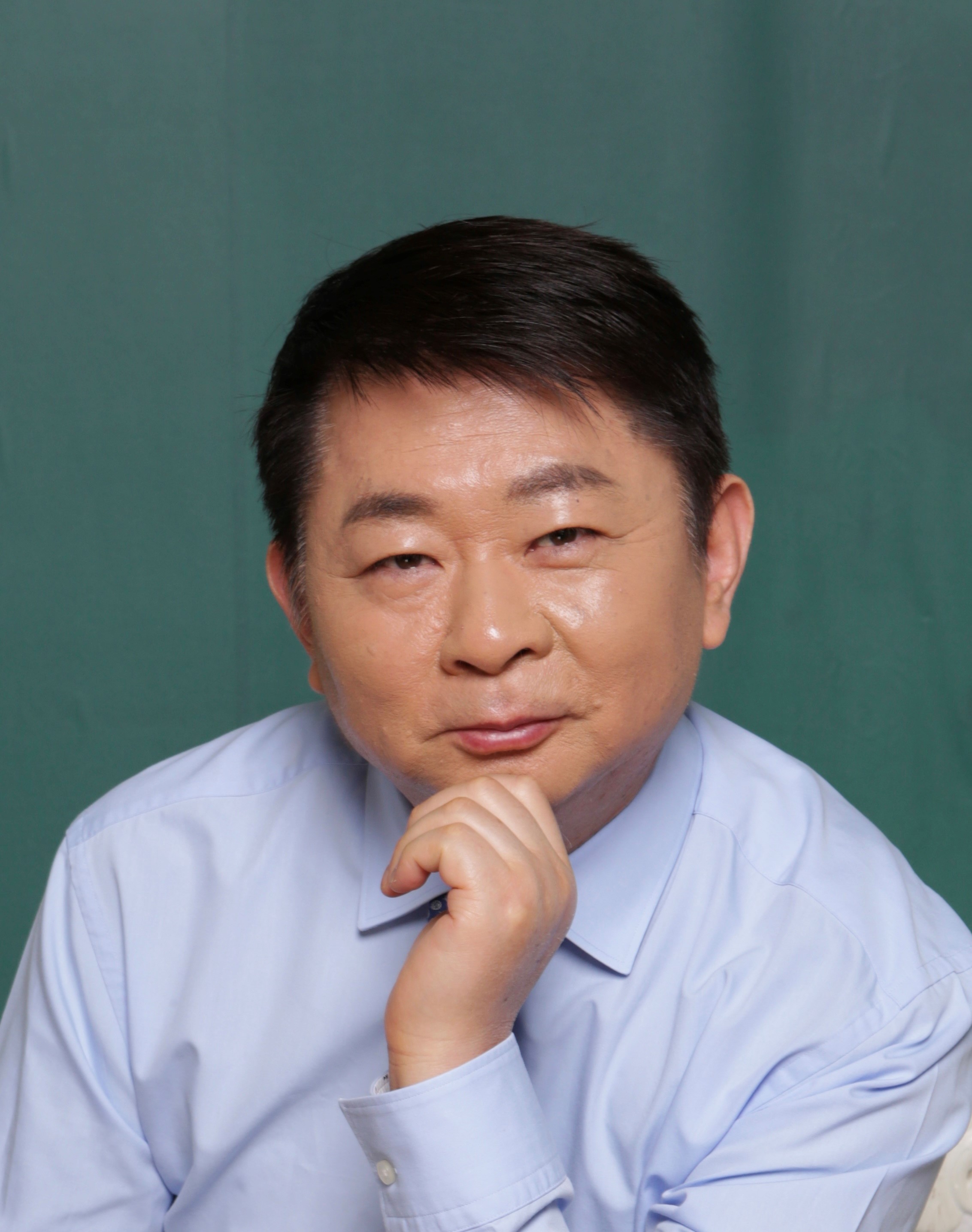}}]{Geoffrey Ye Li}
	(Fellow, IEEE) is currently a Chair Professor at Imperial College London, UK.  Before joining Imperial in 2020, he was a Professor at Georgia Institute of Technology, USA, for 20 years and a Principal Technical Staff Member with AT\&T Labs – Research (previous Bell Labs) in New Jersey, USA, for five years. He made fundamental contributions to orthogonal frequency division multiplexing for wireless communications, established a framework on resource cooperation in wireless networks, and introduced deep learning to communications. In these areas, he has published over 600 journal and conference papers in addition to over 40 granted patents. His publications have been cited over 62,000 times with an H-index of 114. He has been listed as a Highly Cited Researcher by Clarivate/Web of Science almost every year.
	
	Dr. Geoffrey Ye Li was elected to IEEE Fellow and IET Fellow for his contributions to signal processing for wireless communications. He won IEEE Eric E. Sumner Award, IEEE ComSoc Edwin Howard Armstrong Achievement Award, and several awards from IEEE Signal Processing, Vehicular Technology, and Communications Societies.
\end{IEEEbiography}

\end{document}